\documentclass[twocolumn,showpacs,pre,superscriptaddress]{revtex4-1}

\usepackage{amsmath}
\usepackage{tabularx}
\usepackage{graphicx}
\usepackage{color}
\usepackage{times}
\usepackage{ifthen}

\newcommand{\change}[1]{{#1}}

\newcommand{\blue}{}

\graphicspath{{figures/}{data/}}

\newboolean{showall} 
\setboolean{showall}{false}

\begin{document}

\title{Exact joint distributions of three global characteristic
  times for Brownian motion}

\author{Alexander K. Hartmann}
\affiliation{Institut f\"ur Physik, Universit\"at Oldenburg, D-26111
	     Oldenburg, Germany}
\author{Satya N. Majumdar}
\affiliation{LPTMS, CNRS, Univ.  Paris-Sud,  Universit\'e Paris-Saclay,
  91405 Orsay,  France}

\begin{abstract}
We consider three global chracteristic times for a one-dimensional Brownian
motion $x(\tau)$ in the interval $\tau\in [0,t]$: the occupation time $t_{\rm o}$
denoting the cumulative time where $x(\tau)>0$, the time $t_{\rm m}$ at which the
process achieves its global maximum in $[0,t]$ and the last-passage time $t_l$ through
the origin before $t$. All three random variables have the same marginal distribution
given by L\'evy's arcsine law. We compute exactly the pairwise joint distributions
of these three times and show that they are quite different
from each other. The joint distributions display rather rich and nontrivial correlations between these
times. Our analytical results are verified by numerical simulations.

  
\end{abstract}


\maketitle

\section{ Introduction }

Whenever measurements  on physical \cite{fornasini2008},
economical \cite{smith1976}, biological \cite{houle2011} or other 
systems are performed, one is always limited in the amount of
data one can obtain, let it be due to financial, time or fundamental physical 
constraints.
It is then beneficial to have available additional data about
correlations or even the joint distributions involving
at least two relevant quantities, either from previous measurements
or on the basis of modeling. This is of crucial interest if
one quantity is easier to determine
such that costly measurements of the other quantity can be avoided
or reduced.
For disease diagnosis, it is often easy
to access risk factors which allow to restrict expensive or even
dangerous medical examinations to persons with a high disease probability. For
example, osteoporosis \cite{delmas1999} is a very common chronic disease
but has no obvious early symptoms. Still, a significant height loss
is a predictor  for osteoporosis \cite{bennani2009,hillier2012,pluskiewicz2023},
such that expensive
and incriminating X-ray examinations can be restricted only
to high-risk patients.
As another example, the investigation of geological structures
\cite{ragan2012} is of high interest, e.g., to measure the water
content or to find other important resources  below the surface.
These structures can be obtained  most accurately
by invasive methods like core samples obtained from drilling. On the other
hand, seismic measurements allow for a much simpler investigation and correlate
well with results from more expensive methods \cite{sloan2007,yuan2023}.
Similarly, seismic reflection data can be used to predict
promising places for much more costly investigations
by archaeological excavation \cite{hildebrand2007}.

This idea of offsetting the lack of resources in measuring one observable via 
exploiting its correlation with another less expensive observable is of general interest in 
any stochastic time series. For example, the time series may represent the
amplitude of 
earthquakes in a specific seismic region, the amount of yearly rainfall in a given area, the 
temperature records in a given weather station, or
the price of a stock over a given period $t$. While the marginal distributions of single 
quantities can be computed for many time series~\cite{serfozo2009}, there are very few 
results on the joint distributions of two observables associated to a time series. The 
simplest and the most ubiquitous stochastic time series represents the position of a one- 
dimensional Brownian motion (BM) of a given duration $t$, or its discrete-time analogue of a 
one-dimensional random walk~\cite{feller1957,MP2010}, with applications in Physics and 
Chemistry \cite{mazo2008}, Electrical Engineering \cite{doyle1984}, Economics 
\cite{baz2004,hirsa2014} or Social Sciences \cite{schweitzer2007}. This is a paradigmatic 
toy model of a correlated time series where many observables can be computed analytically 
and hence has been extensively used to test various general physical 
ideas~\cite{feller1957,redner2001,majumdar2005,majumdar2010,bray2013}. In this model, the 
joint distribution of local (in time) observables, e.g. of the positions at two different 
times can be easily computed and is simply a bivariate Gaussian distribution. However, 
computing the joint distribution between two global (in time) observables is nontrivial even 
in this simple toy model, since it involves the full history of the trajectory over the 
interval $[0,t]$. For example, the joint distribution of the global maximal value $M(t)$ and 
the time $t_{\rm m}\equiv t_{\max}$ at which it occurs in $[0,t]$ can computed exactly with many different 
applications~\cite{shepp1979,buffet2003,randon-furling2007,MRKY08,MB2008,MCR2010, 
schehr2010,rambeau2011}. Another example consists in computing 
exactly the joint distibution of $t_{\rm  m}$ and $t_{\min}$ (time at which the global minimum 
occurs)~\cite{MMS2019,MMS2020}. This is clearly of interest in finance, because one would 
like to buy a stock at time $t_{\min}$ and sell it at $t_{\max}$.

\begin{figure}
\includegraphics[width = 0.7\linewidth]{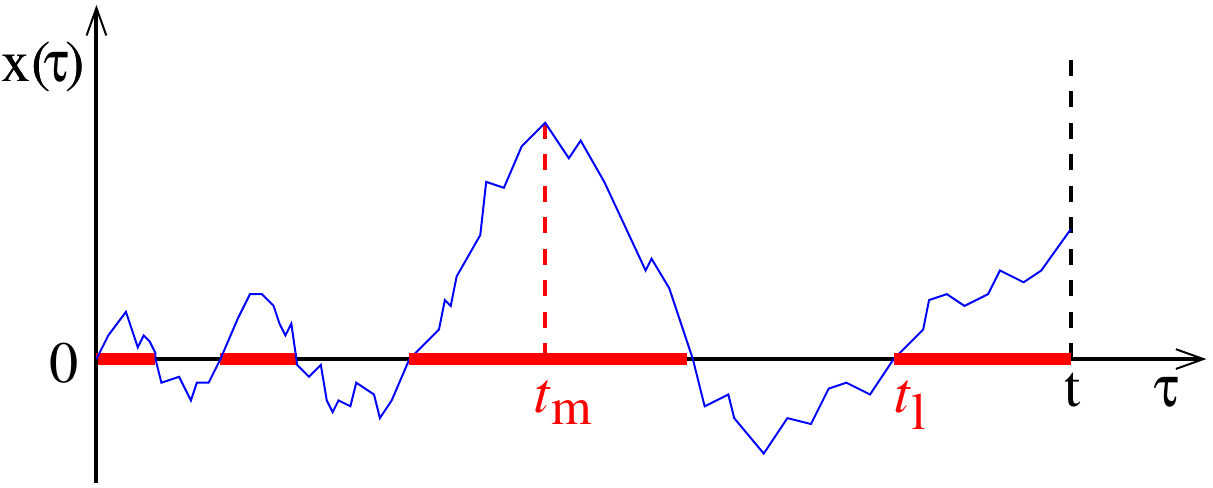}
\caption{A schematic trajectory of a Brownian motion $x(\tau)$ that starts
  at $x(0)=0$ and evolves up to time $t$. Also
  the last-passage time $t_{\rm l}$  and the time of the maximum $t_{\rm m}$ are shown.
The occupation time $t_{\rm o}$ is the sum of the red intervals on the time axis where $x(\tau)>0$.}
\label{fig:bm}
\end{figure}

In this paper, we consider three global quantitites for a time series of duration $t$ starting at the origin.
They are respectively (i) the
occupation time \blue{$t_{\rm o}=\int_0^t \theta(x(\tau))\, d\tau$}, i.e., the cumulative time
where $x(\tau)>0$ with $\theta(z)$ denoting the Heaviside step function: $\theta(z)=1$ for $z>0$
and $\theta(z)=0$ for $z<0$ (ii)
the time $t_{\rm m}$ of the maximum $\max_{\tau} x(\tau)$ in $[0,t]$, and
(iii) the 
time $t_{\rm l}$ of the last passage, before $t$, of the process to $x=0$ (see Fig.~\ref{fig:bm}).
All three observables are random variables each supported over $[0,t]$.
For example, for a time series representing the value of a stock over a maturity period $[0,t]$, normalized
  to the value when bought, the occupation
  time is the time where it exhibits profit,
  the time of the maximum is the best
  time to sell, and the last-passage time indicates the last time the
  value changes from profit to loss or vice-versa.
For a Brownian time series, the marginal distributions
of these random varibles were computed by L\'evy~\cite{levy1940} and quite remarkably,
they all follow the same \emph{arcsine law}
  $p_i(t_i|t)=1/\left[\pi \sqrt{t_i(t-t_i)}\right]$ for $t_i=t_{\rm o}, t_{\rm m}$ or $t_{\rm l}$,
  where the name originates from the corresponding
  arcsine cumulative distribution, i.e., ${\rm Prob.}[t_i\le T|t]=(2/\pi)\, \sin^{-1}\left(\sqrt{T/t}\right)$.

These three quantities are relevant in various complex systems and have been studied 
a for a wide variety of stochastic processes with diverse applications.
  The occupation
  time $t_{\rm o}$ has been analyzed, e.g., for random jump processes~\cite{andersen1954},
renewal and resetting processes~\cite{lamperti1958,godreche2001,burov2011,den_hollander2019}, Gaussian Markov 
processes~\cite{dhar1999,de_smedt2001} and has played
an important role in the context of persistence in nonequilibrium 
systems~\cite{bray2013, newman1998, drouffe1998,toroczkai1999,baldassarri1999}.
Furthermore, it has been studied for various other processes 
such as  opinions in voter models \cite{cox1983}, blinking nanacrystals~\cite{margolin2005}, 
diffusion in a disordered environment~\cite{majumdar2002,sabhapandit2006,radice2020,kay2023},  
a class of spin glass models~\cite{majumdar2002b}, sub- and superdiffusive
processes~\cite{bel2005,barkai2006,delvecchio2024,sadhu2018},
random acceleration process~\cite{boutcheng2016},
active run-and-tumble process~\cite{singh2019,bressloff2020,
mukherjee2023}, permutation genrated random walks~\cite{fang2021} and for a noninteracting Brownian gas~\cite{burenev2024}. 
 Experimentally, the occupation time has been studied for colloidal
  quantum dots \cite{brokmann2003,stefani2009},  thermodynamic currents
  \cite{barato2018}, and  coherently driven
  optical resonators \cite{ramesh2024}.

The time $t_{\rm m}$ at which the global maximum occurs in a process of total duration $t$
has been studied extensively in the context of extreme-value statistics~\cite{majumdar2010,majumdar2020,majumdar2024}.
Examples include
Brownian motion with drift
  \cite{shepp1979,buffet2003,MB2008}, permutation-generated
  random walks \cite{fang2021}, Brownian and anomalous diffusion with constraints
  \cite{MRKY08,randon-furling2007,schehr2010,majumdar2010hitting,MMS21},
  random acceleration
  process \cite{majumdar2010acceleration}, 
  heterogeneous diffusion \cite{singh2022,delvecchio2024},
  run-and-tumble process
  \cite{singh2019,MMS21}, 
  fractional Brownian motion \cite{delorme2016,sadhu2018},
processes with stochastic 
  resetting \cite{MMS21,singh2021}. The lack of symmetry of the distribution of $t_{\rm m}$ around its mean $t/2$
in a stationary time series has been shown to be a sufficient condition that the 
underlying dynamics violates detailed balance~\cite{MMS21,MMS22}.
Another application of $t_m$
of a $1$-d process is to estimate the mean perimeter and the
mean area of the convex hull of a $2$-d process~
  \cite{randon-furling2009,majumdar2010convex,reymbaut2011,dumonteil2013,rtp_ch2020,MMSS21,singh2022convex}.
  Other applications appear in  finance 
  \cite{dale1980,MB2008,chicheportiche2014} and
  sports \cite{clauset2015}.

  The last-passage time $t_{\rm l}$ has been studied also for diffusion in
  extermal potentials \cite{bao2006,comtet2020}, in imhomogeneous environement~\cite{delvecchio2024}, and 
  in permutation-generated
  random walks \cite{fang2021}. For applications, it
  was considered for electronic ring oscillators
  \cite{leung2004,robson2014} and for investigating fission \cite{bao2004}.
  It has been used for setting up Monte Carlo methods
  to calculate capacitances on systems with flat or spherical
  surfaces \cite{hwang2006,yu2021}. The last-passage time has also been
  applied to devise optimal inspection policies
  of degrading systems \cite{barker2009} and in finance~\cite{nikeghbali2013}.

\begin{figure}
\includegraphics[width=0.5\columnwidth]{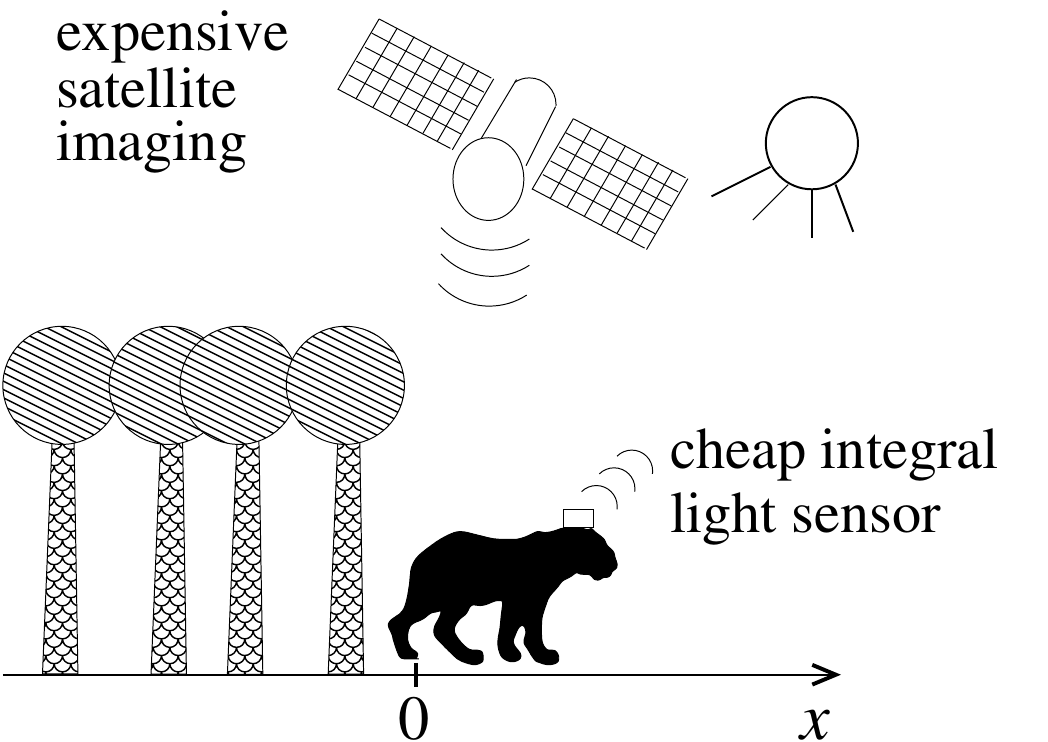}
  \caption{An animal living in a mixed habitat consisting of forests and open
    grass land ($x>0$). Only for $x>0$ it makes sense
    to use expensive stallite imaging. With a low-cost
    and low-weight light sensor
    an estimation of the time $t_{\rm o}$ the animal spends at $x>0$
    allows for better estimate times when to obtain expensive images.
  \label{fig:animal_tracking}}
  \end{figure}

Although the marginal distributions of these three global
  quantities $t_{\rm o}$, $t_{\rm m}$ and  $t_{\rm l}$ have been
  studied for various stochastic time series as mentioned above, their
  mutual  correlations, encoded in pairwise joint distributions, have
  not been studied to the best of our knowledge. As  an example why
  these joint distributions are of interest, one could consider animal
  movement as a toy  application. The development of low-orbiting
  micro satellites like the KITSUNE 6U satellite \cite{azami2022}
  makes the observation with high-temporal imaging possible such that
  in the future even animals can be  observed, but at high cost. For example,
  tigers live in a mixed habitats consisting of deep woods and open
  grass  land \cite{mazak1981}, see
  Fig.~\ref{fig:animal_tracking}. Here, it makes only sense to buy
  expensive high-resolution satellite footings, for times  the animal
  is  in the open land. We assume that one has to decide which
  of the expensive images to buy before being able to look at them.
  Clearly, the best time footage to buy
  is the time $t_{\rm m}$ where the animal is maximally displaced from the wooded
  region so that it is completely visible. However, $t_{\rm m}$ is not
  known apriori
  and has a broad distribution $p(t_{\rm m}|t)$ 
  distributed all over $[0,t]$ (as in the arcsine law for a
  one-dimensional Brownain motion). So, it is not clear which
  time or times to choose.  However, suppose we
  have some knowledge about the occupation time $t_{\rm o}$ that
  corresponds to the cumulative time an animal spends in the open
  land. This could be measured easily
  with a low-cost and low-weight  sensor which
  does not annoy the animal considerably, i.e., without relatively
  heavy GPS or time-stamping  facilities.  Conditioned
  on the knowledge of $t_{\rm o}$, a representative value of $t_{\rm
    m}$ with high probability may emerge and one can then buy the
  footage corresponding to that. We will show an explicit example of
  that for the $1$-d Brownian motion later (in the right panel of
  Fig.~\ref{fig:xoxm}).  Thus, exploiting the correlations between two
  global quantities in a model, here for a simple random walk, allows
  for a much better targeted analysis of the data. 

To gain better insights into the correlations between these
  three global quantities,  one needs to compute the
  pairwise joint distributions of them analytically in a simple model
  time series. \blue{We have not seen 
any result in the literature on the joint distributons of these three global observables
for any stochastic process, including even for the simple one-dimensional
Brownian motion. This lack of benchmark results motivated us to study the joint bivariate
distributions between these three observables for a
one-dimensional Brownian motion
of duration $t$, perhaps the simplest and most ubiquitous one-dimensional stochastic 
process. For this model, the marginal distributions of all
  three quantities are identically given by the arcsine law, as
  mentioned earlier. Here we compute exactly the pairwise joint
  distributions and demonstrate that, even in this simple model, they
  exhibit a rather rich, nontrivial, and unexpected behavior, not at
  all compatible with, say, standard linear correlations. We hope that the
results and the methods presented in the paper will mark the starting
point for computing and exploring these joint distributions for other stochastic processes as well.}

\blue{The rest of the paper is organised as follows. In Section 
\ref{model} we introduce the model and the three observables of interest more
precisely. In Section \ref{results}, we provide a brief outline
of the derivations and also summarize our main results.  
Section \ref{conclusion} contains a conclusion and discussion. Since the actually derivations
of these results are somewhat technical, we relegate them to 
four Appendices.}


\section{Model and observables of interest}
\label{model}

\blue{The model stochastic process we study in this paper is perhaps the simplest,
namely
a one \blue{dimensional} Brownian motion of duration $t$ evolving via 
the Langevin equation
\begin{equation}
\frac{dx}{d\tau}= \sqrt{2D}\, \eta(\tau)\, ,
\label{Lange.1}
\end{equation}
where $D$ is the diffusion constant and $\eta(\tau)$ is a Gaussian white 
noise with zero mean and delta correlator: $\langle
\eta(\tau_1)\eta(\tau_2)\rangle=\delta(\tau_1-\tau_2)$. We assume that the process 
starts at the origin, i.e., $x(0)=0$.}

\blue{ The three classical observables of interest are last-passage time, occupation time
and time of maximum, already introduced briefly in the introduction. Below
we define them more precisely (see Fig. (\ref{fig:bm}) for a schematic representation).}

\begin{itemize}

\item {\bf Last-passage time.} \blue{Let $t_{\rm l}$ denote that last time before
$t$ that the Brownian motion crosses the origin (from either side). Evidently $0\le t_{\rm l}\le t$.}

\item {\bf Occupation time.} \blue{This is the total time that
the Brownian motion of duration $t$ spends
on the positive half axis, i.e.,
\begin{equation}
t_{\rm o}= \int_0^t \theta\left(x(\tau)\right)\, d\tau\, ,
\label{occup.1}
 \end{equation}
where $\theta(x)$ is the standard Heaviside theta function:
$\theta(x)=1$ for $x>0$ and $\theta(x)=0$ for $x<0$.
Clearly $0\le t_{\rm o}\le t$.}

\item{\bf Time of the maximum.} \blue{We denote by $t_{\rm m}$ the time at which
the Brownian motion achieves
its maximum (on the positive side) within the interval $[0,t]$.
Again, $0\le t_{\rm m}\le t$.}

\end{itemize}

\blue{
These three observables are each supported over
$[0,t]$ and they are random variables since they fluctuate from sample
to another. The marginal distributions of these random variables were
computed by L\'evy~\cite{levy1940} and it is well known that they all share the
same probability distribution function (PDF). In fact, the three
marginal PDF's, given the duration $t$, can be written in the scaling
form 
\begin{equation}
p_i\left(t_i|t\right) =
\frac{1}{t}\, f\left(\frac{t_i}{t}\right)\, ,\quad\quad\quad
{\rm with}\quad 0\le t_i\le t\, ,
\label{eq:arcsine.pdf}
\end{equation}
valid for all $t$,
where $t_i=t_{\rm o}$, $t_{\rm m}$ or $t_{\rm l}$. The scaling function
$f(x)$ is identical for all three observables and is given explicitly by
\begin{equation}
f(x)= \frac{1}{\pi\, \sqrt{x\,(1-x)}}\, \quad \quad
{\rm with}\quad 0\le x\le 1\, .
\label{fx_def}
\end{equation}
Interestingly, the marginal distributions of all three observables
$t_{\rm l}$, $t_{\rm o}$ and $t_{\rm m}$ are independent
of the diffusion constant $D$ of the Brownian motion.
The cumulative distribution of each of them has the arcsine form:
${\rm Prob.}[x\le u]= \int_0^u f(x)\, dx=(2/\pi)\,
\sin^{-1}\left[\sqrt{u}\right]$. Hence, these three laws are famously known as
arcsine laws of L\'evy~\cite{levy1940}.}

\blue{One interesting fact about the marginal distribution $f(x)$ in 
Eq. (\ref{fx_def}) is that it has a ${\rm U}$ shape with peaks at the two end points $x=0$
and $x=1$. In other words, the typical values of $x$ (where the PDF achieves its maximum value)
do not coincide with the average $\langle x\rangle= \int_0^1 x\, f(x)\, dx=1/2$. This
is very different from the bell shaped curves whose peak coincides with its mean.
For example, consider the observable $t_{\rm o}$ denoting the occupation time.
Its scaled value $x=1$ means $t_{\rm o}=t$, i.e., it correesponds to paths
that, starting at the origin, stay entirely on the positive side during $[0,t]$.
On the other hand $x=0$, i.e., $t_{\rm o}=0$ corresponds to paths that, starting at
the origin, stay entirely on the negative side during $[0,t]$.
Thus, physically this ${\rm U}$ shaped curves indicate that the typical trajectories of
a Brownian motion are `stiff', i.e., if they deviate to the positive (negative) side, they
tend to stay positive (negative) for the full interval $[0,t]$. These facts are, by no means,
very intuitive even for this simple Brownain motion.}

\blue{Furthermore, it is also not obvious at all why these three observables should have the same marginal
distributions. While there exist probablistic equivalences between these observables
using some special properties of Brownian motion~\cite{levy1940}, for a physicist
it is not easy to follow these mathematical equivalences. The arcsine laws for these
three observables have been derived in the 
physics literature using  methods that are quite different for each observable.
For example, the arcsine law for the occupation time $t_{\rm o}$ can be derived by
using the Feynman-Kac method involving path-integrals~\cite{BF2005}.
On the other hand, the derivations of the arcsine laws for $t_{\rm m}$ and
$t_{\rm l}$ involve path decomposition methods that involve spilitting the
full trajectory into two parts: (i) the interval $[0, t_{\rm m}]$ prior to $t_{\rm m}$ (or for $t_{\rm l}$)
and (ii) the interval $[t_{\rm m}, t]$ after the occurrence of $t_{\rm m}$ (and similarly for $t_{\rm l}$).
The probability weight of each part is then computed separately and then they are multipled (using the Markov property
of the process) to get the full probability distribution of the observable $t_{\rm m}$
and $t_{\rm l}$. For a derivation of the arcsine law for $t_{\rm m}$ using this path decomposition method, see
Ref.~\cite{MRKY08}. We also rederive these marginal laws later in the Appendices from our more
general results on joint bivariate distributions of these three variables.}

\blue{While the marginal PDF's of these fractions are identical, 
our goal here is to compute their bivariate joint distributions}
\begin{eqnarray}
{\rm Prob.}\left[t_{\rm l} ,\,  t_{\rm o}\right|t] & =& P_{12}(t_{\rm l},\,t_{\rm o}|t)
\, , \nonumber\\
{\rm Prob.}\left[t_{\rm o} ,\, t_{\rm m}|t\right] & = &
\change{P_{23}(t_{\rm o},\,t_{\rm m}|t)}\, , \\
{\rm Prob.}\left[t_{\rm l} ,\, t_{\rm m}|t\right] & = & P_{13}(t_{\rm l},\,t_{\rm m}|t)\, .
\nonumber 
\label{P_def}
\end{eqnarray}
The subscripts $1$, $2$ and $3$ refer respectively to
$t_{\rm l}$, $t_{\rm o}$ and $t_{\rm m}$.
\blue{On dimensional grounds, these three joint distributions can be expressed in the scaling form, valid for all $t$,
\begin{eqnarray}
P_{12}(t_{\rm l},\,t_{\rm o}|t) & = &\frac{1}{t^2}\, {\cal P}_{12}\left(\frac{t_{\rm l}}{t},\, \frac{t_{\rm o}}{t}\right)
\, , \nonumber\\
P_{23}(t_{\rm o},\,t_{\rm m}|t) &= & \frac{1}{t^2}\, {\cal P}_{23}\left(\frac{t_{\rm o}}{t},\, \frac{t_{\rm m}}{t}\right)
    \, , \nonumber \\
P_{13}(t_{\rm l},\,t_{\rm m}|t)  & = & \frac{1}{t^2}\, {\cal P}_{13}\left(\frac{t_{\rm l}}{t},\, 
\frac{t_{\rm m}}{t}\right)\, .
\label{P_scale_def}
\end{eqnarray}
We will show that these three joint distributions are highly 
nontrivial and are rather different from each other.
We compute them exactly using the $\epsilon$-path decomposition method, first
proposed in Refs.~\cite{MC2004,MC2005} and then used subsequently 
in numerous other related
problems~\cite{majumdar2005,MRKY08,PCMS13,MMS2019,MMS2020,MMS21,MMSS21,MMS22}.
Below we outline the main ideas leading to the results
and the details of the derivations are presented
in the Appendices since they are a bit technical.}


\section{Results}
\label{results}

\blue{In this Section, we present the main results for the joint bivariate distributions of the three
main observables and discuss their physical implications. We also briefly outline here the main ideas
behind the derivations of these results, while leaving the technical details
to the Appendices.}

\subsection{Last-passage time $t_{\rm l}$ and occupation time $t_{\rm o}$} 

First, we focus on computing the
$P_{12}(t_{\rm l},\,t_{\rm o}|t)$, because this is the simplest of the three. 
The idea is to first express the 
joint distribution as
$P_{12}(t_{\rm l},t_{\rm o}|t)= P_{\rm cond}(t_{\rm o}|t_{\rm l},t)\, p_1(t_{\rm l}|t)$
where $P_{\rm cond}(t_{\rm o}|t_{\rm l},t)$ denotes the conditional
distribution of $t_{\rm o}$ given $t_{\rm l}$ and $t$, while $p_1(t_{\rm l}|t)$ is the
marginal distribution of $t_{\rm l}$ given in Eq. (\ref{eq:arcsine.pdf}). 
We split the total duration $t$ into two intervals:
(${\rm I}$) $0\le \tau \le t_{\rm l}$ and (${\rm II}$) 
$t_{\rm l}\le \tau\le t$.
Consequently, $t_{\rm o}$ can also be split into two parts
$t_{\rm o}= t_{\rm o}^{\rm I} + t_{\rm o}^{\rm II}$, where the two random
variables $t_{\rm o}^{\rm I}$ and $t_{\rm o}^{\rm II}$ are independent
of each other, for a given fixed $t_{\rm l}$.
The idea then is to compute
the PDF of each of them separately and then 
convolute them to compute the full conditional probability.

For the first interval,
the Brownian motion starts
at the origin at $\tau=0$ and ends at the origin at $\tau=t_{\rm l}$. 
\blue{This then corresponds to a \emph{Brownian bridge} of duration $t_{\rm l}$, starting
and ending at the origin.
We then need to compute the distribution of the
occupation time $t_{\rm o}^{\rm I}$ of this Brownian bridge of duration $t_{\rm l}$.
In fact, the occupation time for a Brownian bridge is a well studied object~\cite{feller1957}
and it is well known that
the PDF of the occupation time $t_{\rm o}^{\rm I}$ is flat, i.e.,
$p_{\rm I}(t_{\rm o}^{\rm I}|t_{\rm l})= \frac{1}{t_{\rm l}}\, \theta\left(t_{\rm l}- 
t_{\rm o}^{\rm I}\right)$. This result can be understood intuitively as follows. Since a
\mbox{Brownian} bridge will return to the origin after period $t_{\rm l}$, one can take any time instant $\tau_0\in  [0,t_{\rm l}]$ and
consider the trajectory over $[\tau_0, \tau_0+t_{\rm l}]$ with an imposed
periodicity $x(\tau)= x(\tau-t_{\rm l})$ for $\tau>t_{\rm l}$.
Subtracting $x(\tau_0)$ so that this new trajectory
also goes from from origin to origin in time $t_{\rm l}$, one has a new Brownian bridge configuration which has
the same weight as the original one. However, the occupation time in this new configuration
is quite different from the original one. Thus, one can generate any value of the occupation number
by sliding and stiching the original trajectory appropriately, but all with equal weights. This leads
to the fact all allowed values of occupation time are equally likely for a Brownian bridge.}

\begin{figure}[ht]
\includegraphics[width = 0.49\linewidth]{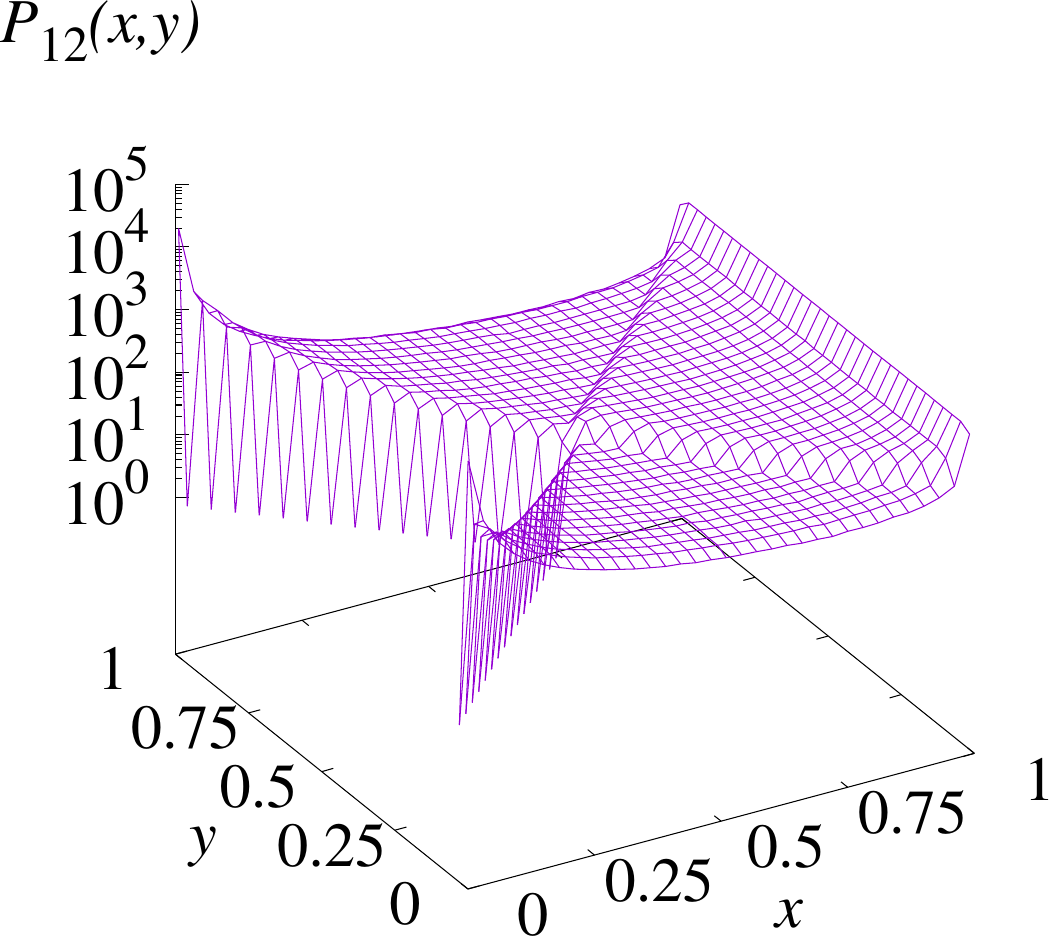}
\includegraphics[width = 0.49\linewidth]{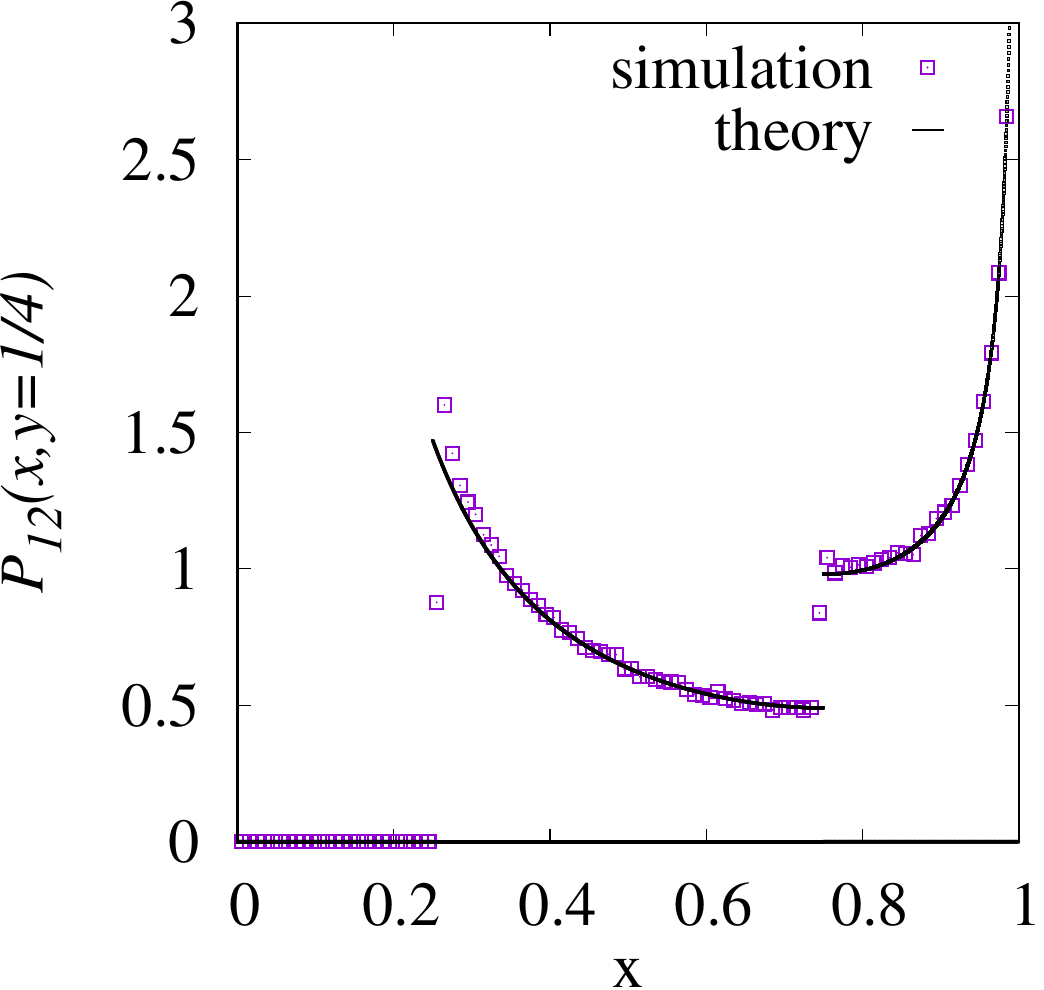}
\caption{(left) The scaled joint PDF ${\cal P}_{12}(x,y)$ of the fractions 
$x=t_{\rm l}/t$ and $y=t_{\rm o}/t$. (right) The distribution
  with fixed $y=1/4$.
The simulation results are shown by open symbols.
\label{fig:xlxo}}
\end{figure}

Concerning the second interval, the process crosses zero for the last time at
$\tau=t_{\rm l}$,
it may either cross from above or from below, each
with equal probability $1/2$.
For the first case, the process stays below zero, so $t_{\rm o}^{\rm II}=0$,
in the other case we have $t_{\rm o}^{\rm II}=(t-t_{\rm l})$. 
Thus, we obtain
$ p_{\rm II}(t_{\rm o}^{\rm II}|t_{\rm l})= \frac{1}{2}\, \delta\left(t_{\rm o}^{\rm II}\right)
+\frac{1}{2}\, \delta\left(t_{\rm o}^{\rm II}- (t-t_{\rm l})\right)$.
The distribution for the  total occupation time $t_{\rm o}=t_{\rm o}^{\rm I}+t_{\rm o}^{\rm II}$
is given by the convolution
$P_{\rm cond}(t_{\rm o}|t_{\rm l},t)=  \int_0^t dt_{\rm o}^{\rm I}\, p_{\rm I}(t_{\rm o}^{\rm I})\,
p_{\rm II}(t_{\rm o}-t_{\rm o}^{\rm I})$.
\blue{We then apply the arcsine result for the marginal distribution of $t_{\rm l}$ from
Eq. (\ref{eq:arcsine.pdf}) and obtain a scaling form (see Appendix \ref{sec:tl:to} for details)
\begin{equation}
P_{12}(t_{\rm l},t_{\rm o}|t)=
\frac{1}{t^2}\, {\cal P}_{12}\left(\frac{t_{\rm l}}{t}=x,\, \frac{t_{\rm o}}{t}=y\right)\, ,
\label{scaling_form_12}
\end{equation}
where the scaling function is given by}
\begin{eqnarray}
{\cal P}_{12}(x,y)& =& \frac{1}{2\,\pi\, x^{3/2}\, 
  \sqrt{1-x}}\left[\theta(x-y)+\theta(x+y-1)\right]\, \nonumber \\
& & \times \, \theta(1-x)\, \theta(1-y)\, .
\label{tlto_scaled.1}
\end{eqnarray} 
\blue{The scaling function is plotted in Fig.~\ref{fig:xlxo}.}
To obtain a better impression
of the distribution, a cross section at $y=1/4$ is also displayed on the right panel of Fig.~\ref{fig:xlxo}.

\blue{Some features of the result for ${\cal P}_{12}(x,y)$ in Eq. (\ref{tlto_scaled.1}) can
be guessed/understood physically. For example, we see from Fig. \ref{fig:xlxo} (both in
the left and in the right panel) that ${\cal P}_{12}(x,y)$ 
vanishes unless the scaled last-passage time $x>y$ or $x>1-y$, i.e., $x>{\min}(y, 1-y)$
for a given fixed
scaled occupation time $y$. In other words the scaling function ${\cal P}_{12}(x,y)$
vanishes if $x< {\min}(y, 1-y)$. For example, in the right panel of figure~\ref{fig:xlxo}
where $y=1/4$, we see that the joint PDF vanishes for $x<y=1/4$. 
One can associate a physical picture to this result
as follows. Consider the event when the last-passage through the origin 
before $t$ where the trajectory crosses from the positive to the negative side.
In this case the occupation time contribution from the interval $[t_{\rm l}, t]$
is zero since the trajectory can not recross back to the positive side in $[t_{\rm l}, t]$.
Thus the occupation time has contribution only from the interval $[0,t_{\rm l}]$
and consequently $t_{\rm o}< t_{\rm l}$, i.e., $x>y$. In other words, for a given $x$,
we must have $y<x$ for ${\cal P}_{12}(x,y)$ to be nonzero.
On the other hand, if the last-passage through the origin occurs from the negative
to the positive side, the walker will spend the remaining time $t-t_{\rm l}$
 at positive positions, and might even have spent some time before
 the last-passage in the positive range. Thus, the occupation time
 must be equal to or larger than $t-t_{\rm l}$, i.e., $y>1-x$.
Hence, we must have $y>1-x$ for ${\cal P}_{12}(x,y)$ to be nonvanishing.
Combining these two events, we then conclude (without doing any computation) that
${\cal P}_{12}(x,y)$ must be nonzero if and only if $y$ does not belong
to the interval $[x,1-x]$, or equivalently 
if $x> {\min}(y, 1-y)$. Let us remark that this argument is 
rather general, and is completely
independent of the actual random process.} 

\blue{Of course,
this argument does not provide us the amount of jump discontinuities at $x=y$ or $x=1-y$ as seen, e.g.,
in the right panel of Fig. (\ref{fig:xlxo}). The magnitudes of such
discontinuities in the joint PDF do depend on the specific process and
can be read off the exact formula in Eq. (\ref{tlto_scaled.1}) for a Brownian motion. Also,
for fixed $y$, we see from Eq. (\ref{tlto_scaled.1}) that 
the scaling function diverges as ${\cal P}_{12}(x,y)\sim (1-x)^{-1/2}$ as $x\to 1$ (see also the right panel
of Fig. (\ref{fig:xlxo})). This divergence as $x\to 1$ typically corresponds to paths that stay almost positive for their
entire duration $t$, except for a little excursion to the origin and re-crossing it from below
at time $t_{\rm l}$ for the last time. Also, notice that
unlike the marginal arcsine law of $t_{\rm o}$, the joint PDF of $t_{\rm o}$, conditioned on a fixed
$t_{\rm l}$, is not symmetric around its average value $\langle t_{\rm o}\rangle =t/2$. 
Let us make a cautionary remark here
by emphasizing that all these little nice features of the joint PDF
may be understood heuristicallly once we have the exact formula as in Eq. (\ref{tlto_scaled.1}).
But without such an exact formula, one will not be able to make such heuristic guesses very confidently. }

Here, also a comparison with numerical
simulations of random walks
is included. For that purpose, standard random walks, starting at zero,
 are performed with
a certain number $n_t$ of  steps. Thus,
the position at step $n$
is given \blue{by $x(n)=\sum_1^n \eta(n)$, where $\eta(n)$'s are independently
  and identically distributed standard Gaussian increments},
numerically
drawn by using the Box-Muller approach \cite{parctical_guide2015}.
The data can be scaled to any time span $[0,t]$ simply by rescaling the
times to $nt/n_t$. This approximates Brownian motion with step size
$\Delta t = t/n_t$. Here, we use $n_t=10000$
and average over independent $10^8$ walks.
Note that only the signs of the positions $x(n)$
are important here, so no position rescaling is necessary.
All three times 
$t_{\rm o}$, $t_{\rm l}$ and $t_{\rm m}$ can be directly read off from the time series
$x(n)$ and used to create corresponding joint histograms.
The accuracy is limited by the discretised sampling at
resolution $\Delta t$, but this
is sufficient for the present study, as visible in Fig.~\ref{fig:xlxo}.
For even better accuracy of $t_{\rm l}$ and $t_{\rm m}$, one could use adaptive interval
splitting approaches \cite{walter2022}.

Clearly, the joint distribution in \eqref{tlto_scaled.1} does not factorize demonstrating nontrivial
  correlations between the scaled variables $x=t_{\rm l}/t$ and $y=t_{\rm o}/t$,
  which we computed explicitly. Interestingly,
while the first-order correlation vanishes $\langle xy\rangle-\langle x\rangle \langle y\rangle=0$, higher
order correlations are nonzero. For example, we find that $x$ and $y$ are
weakly anticorrelated
$\langle  x^2 y^2\rangle-\langle x^2\rangle \langle y^2\rangle= - 7/{384}$,
see Eq.~(\ref{cov22.1}) in \blue{Appendix \ref{sec:tl:to}, a result that is also confirmed
numerically.}

\blue{The first order correlation is zero, because for any path $\{x[n]\}$
  in $[0,t]$,
which has last-passage time $t_{\rm l}$ and occupation time $t_{\rm o}$,
one has the path $\{-x[n]\}$ which has exactly
the same last-passage time $t_{\rm l}$
and the 'opposite' occupation time $t-t_{\rm o}$. Since both paths have the
same probability, the dependence between $t_{\rm l}$ and $t_{\rm o}$
averages out to zero.
The observed higher-order anticorrelation between $t_{\rm o}$ and $t_{\rm l}$ has
the following physical
implication. It indicates that when $t_{\rm o}$ is large,
typically $t_{\rm l}$ is small
and vice versa. Now $t_{\rm o}$ large indicates paths that stay on the
positive side for
a long time and they seem to cross the origin quite early so that $t_{\rm l}$
is small.
On the other hand, if $t_{\rm o}$ is small, the typical path stays on the negative side  for a long time
and crosses zero from the negative to the positive side only near the end of the time interval $[0,t]$,
indicating that $t_{\rm l}$ is relatively large.  Let us emphasize that
such weak higher-order
anticorrelations between $t_{\rm l}$ and $t_{\rm o}$ could not have been 
guessed without the explicit exact result in Eq. (\ref{tlto_scaled.1}).}


\subsection{Occupation time $t_{\rm o}$ and time of maximum $t_{\rm m}$} 

\blue{Computing the joint distribution
$P_{23}(t_{\rm o},t_{\rm m}|t)$ turns out to be harder than computing ${\cal P}_{12}(t_{\rm l}, t_{\rm o}|t)$
in the previous subsection. Here the main idea is to first compute a more complicated
object $P(t_{\rm o}, t_{\rm m}, M|t)$ denoting the joint distribution of $t_{\rm o}$, $t_{\rm m}$
and the value of the maximum $M$ within the interval $[0,t]$. Once we know this, the joint
distribution $P_{23}(t_{\rm o},t_{\rm m}|t)$ can be computed by marginalising over $M$, i.e.,
\begin{equation}
P_{23}(t_{\rm o}, t_{\rm m}|t)= \int_0^{\infty} P(t_{\rm o}, t_{\rm m}, M|t)\, dM\, .
\label{marginal_23}
\end{equation}   
Now, this quantity $P(t_{\rm o}, t_{\rm m}, M|t)$ can be computed again by splitting the
full trajectory into two parts: (i) left part over $[0,t_{\rm m}]$ where the trajectory
propagates from $x(0)=0$ to $x(t_{\rm m})=M$ at time $t_{\rm m}$ and (ii) right part over $[t_{\rm m}, t]$
where the trajectory propagates from $x(t_{\rm m})=M$ at time $t_{\rm m}$ to $x(t)$ at time $t$ where
the final position will be integrated over $[-\infty, M]$.
The total occupation time $t_{\rm o}$ is again the sum of the occupation times
of the left and the right intervals. The distribution of $t_{\rm o}$ is
then the convolution of the possible values
of $t_{\rm o}$ for the left and the right interval. Thus, one can formally write down
$P(t_{\rm o}, t_{\rm m}, M|t)$ by multiplying the appropriate propagators of the left and the right
intervals by adapting the Feynman-Kac method. The full computation eventually gives quite powerful and nontrivial
result, but the derivation is rather technical and hence we present the details 
in Appendix \ref{sec:to:tm}. The upshot at the end of the day is that this method
gives us an exact formula for the triple Laplace transform of $P_{23}(t_{\rm o}, t_{\rm m}|t)$
with respect to $t_{\rm o}$, $t_{\rm t}$ and $t$, as given in Eq. (\ref{tlpt.1}) in
Appendix \ref{sec:to:tm}. 
Unfortunately, this triple Laplace
transform is very hard to invert explicitly. Consequently, unlike
the scaled joint distributions ${\cal P}_{12}(x,y)$ and ${\cal P}_{13}(x,y)$,
it is hard to obtain the scaled joint distribution ${\cal P}_{23}(x,y)$
explicitly. However, as we will see in Appendix \ref{sec:to:tm}, the
triple Laplace transform does give explicit access to correlation moments
between $t_{\rm o}$ and $t_{\rm m}$.}

\begin{figure}
  \begin{center}
    \includegraphics[width = 0.49\linewidth]{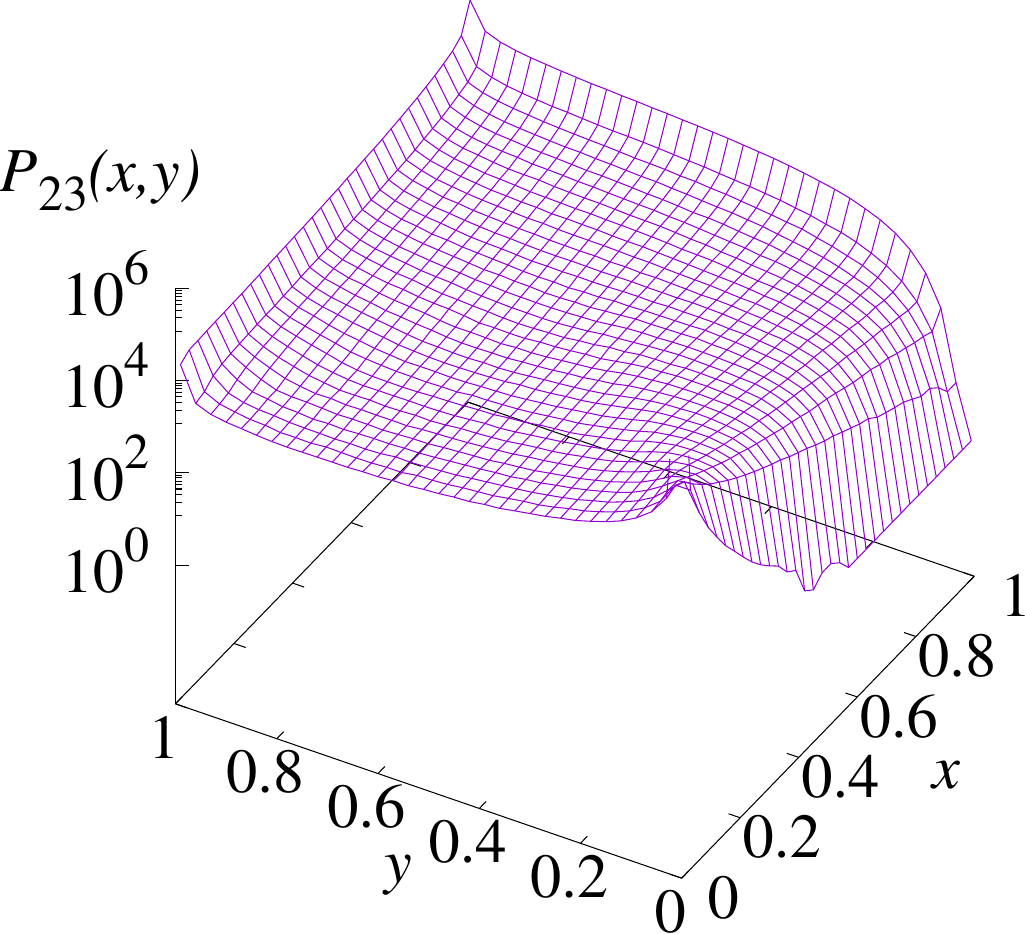}
\includegraphics[width=0.49\columnwidth]{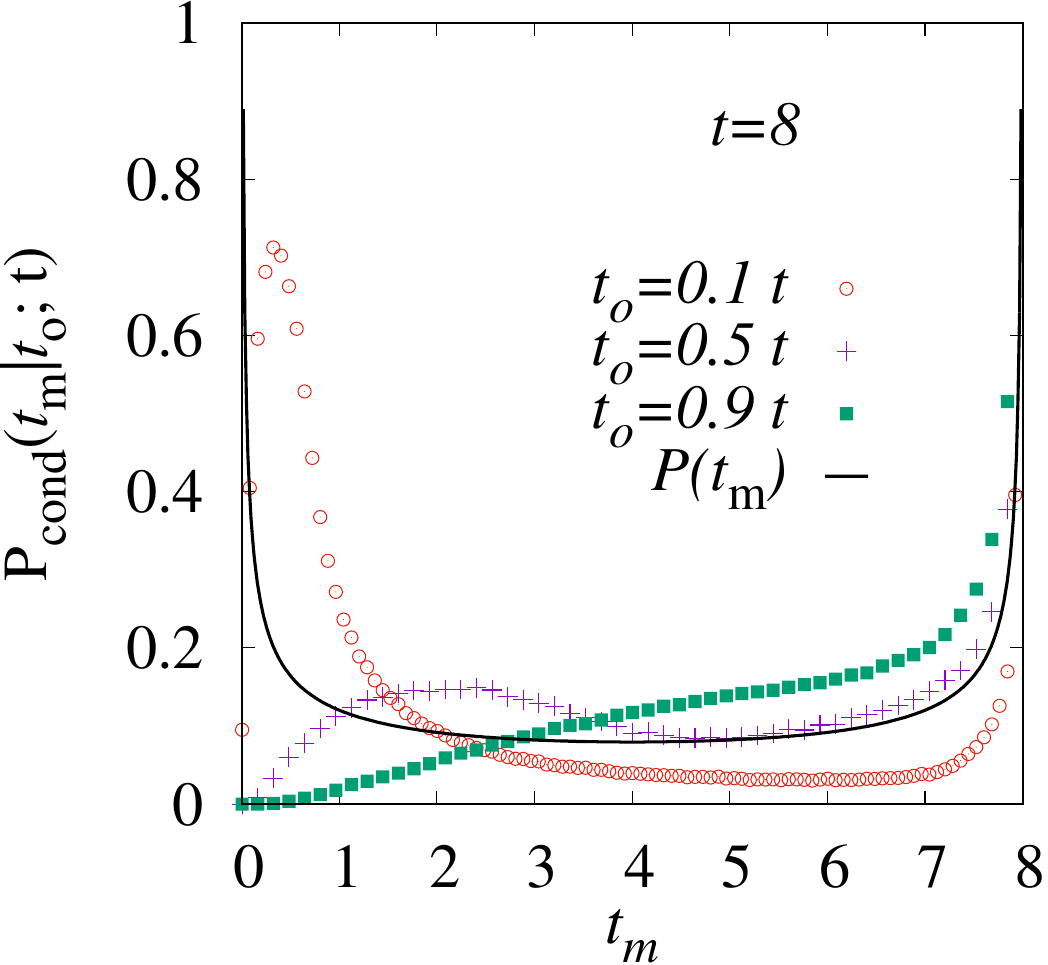}
  \end{center}
\caption{(left) The scaled joint PDF ${\cal P}_{23}(x,y)$ of the fractions 
$x=t_{\rm o}/t$ and $y=t_{\rm m}/t$ plotted in the $(x-y)$ plane. 
(right) Corresponding distribution of the time $t_{\rm m}\in [0,t]$ (here $t=8)$
  of the maximum position,
  conditioned to different values of the occupation time $t_{\rm o}$.
  \label{fig:xoxm}}
\end{figure}

\blue{Indeed, from Eq. (\ref{tlpt.1}) in Appendix \ref{sec:to:tm}, 
we find that the two marginal distributions for $t_{\rm o}$ and $t_{\rm m}$ result in
the arcsine laws, and also obtain various covariance functions
explicitly. For example, we get 
\begin{eqnarray}
\label{cov:to:tm}
\langle t_{\rm o}\, t_{\rm m}\rangle-
\langle t_{\rm o}\rangle\,\langle t_{\rm m}\rangle  &= & \frac{11}{144}\, t^2 \nonumber \\ 
\langle t_{\rm o}\, t_{\rm m}^2\rangle-\langle t_{\rm o}\rangle\, 
\langle t_{\rm m}^2\rangle &= & \frac{43}{648}\, t^3 \nonumber \\
\langle t_{\rm o}^2\, t_{\rm m}\rangle-
\langle t_{\rm o}^2\rangle\, \langle t_{\rm m}\rangle &=&
\frac{509}{7200}\, t^3 \nonumber \\
\langle t_{\rm o}^2\, t_{\rm m}^2\rangle-
\langle t_{\rm o}^2\rangle\, \langle t_{\rm m}^2\rangle 
&= &  \frac{25997}{414720}\, t^4 \, ,  
\end{eqnarray}
which are all compatible with our numerical results. The explicit results in \eqref{cov:to:tm}
indicate that $t_{\rm o}$ and $t_{\rm m}$ are actually positively correlated, in contrast to
the pair $(t_{\rm o}, t_{\rm l})$ which are anticorrelated. The positive correlation implies that if  
$t_{\rm o}$ is large then $t_{\rm m}$ is likely to be large and vice versa. This has the following physical
implications. Consider a trajectory that has a large $t_{\rm o}$, i.e., it stays on the positive side
for a long time. Then, for such a trajectory, the maximum is likely to occur at the end of the interval
$[0,t]$ implying $t_{\rm m}$ to be relatively large. Similarly, if $t_{\rm o}$ is small, it corresponds
to trajectories that spend most of their lifetime on the negative side. The positive correlations between
$t_{\rm o}$ and $t_{\rm m}$ indiacte that for such trajectories the maximum will likely occur
at earlier times such that $t_{\rm m}$ is small. Those paths start from the origin, go upwards
to achieve their maximum at an early time with $t_{\rm m}$ small and then turn around, cross zero and stay negative
till the final time $t$.}

In Fig.~\ref{fig:xoxm}, the scaled joint PDF
${\cal P}_{23}(x,y)$ obtained from the numerical simulations is shown.
Here, no discontinuities are present, but for fixed value $x=t_{\rm o}/t$
of the scaled occupation time, the behavior of the probability
changes considerably.
For small scaled occupation times $x$, the
distribution is large for small times $y$ of the time of maximum, while
for large occupation times $x\approx 1$ it is opposite. This is also
visible in the distribution of $t_{\rm m}$ conditioned to a value of $t_{\rm o}$,
as shown in the right panel of Fig.~\ref{fig:xoxm}.

\begin{figure}[ht]
\includegraphics[width = 0.7\linewidth]{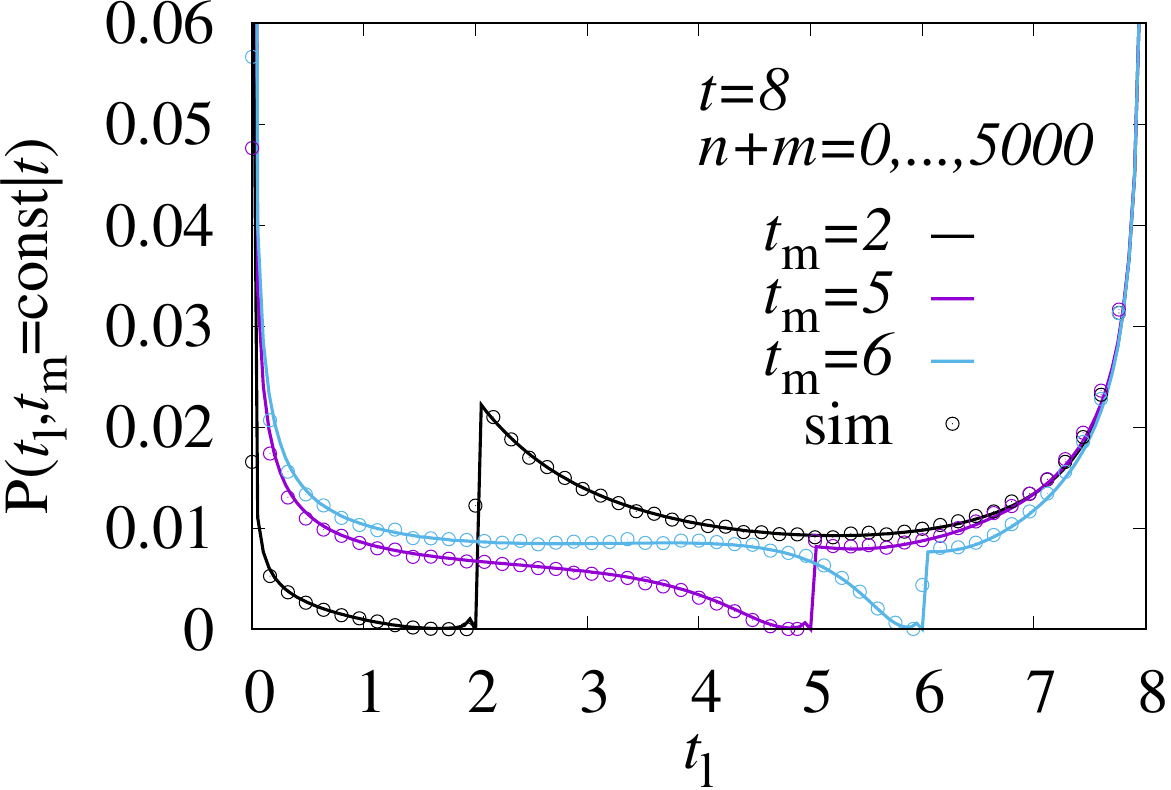}
\caption{The joint PDF $P_{13}(t_{\rm l},t_{\rm m}|t)$
  from numerically evaluating Eq. (\ref{jointP13.1}) with $t=8$,
  shown as a function of
$t_{\rm l}$ for three different fixed
values of $t_{\rm m}$. Simulation results are shown as symbols.
\label{fig:tltm}}
\end{figure}


\subsection{Last-passage time $t_{\rm l}$ and time of maximum $t_{\rm m}$} 

\blue{In this subsection, we compute the joint bivariate distribution
$P_{13}(t_{\rm l}, t_{\rm m}|t)$.
As in the previous subsection, it turns out to be convenient to first consider
a more complicated object, namely the joint distribution $P(t_{\rm l}, t_{\rm m}, M|t)$
of $t_{\rm l}$, $t_{\rm m}$ and the value $M$ of the maximum up to time $t$.
Upon obtaining this three variable joint distribution, one can obtain
$P_{13}(t_{\rm l}, t_{\rm m}|t)$ via the marginalisation
\begin{equation}
P_{13}(t_{\rm l}, t_{\rm m}|t)= \int_0^{\infty} P(t_{\rm l}, t_{\rm m}, M|t)\, dM\, .
\label{marginal_13}
\end{equation} 
The quantity $P(t_{\rm l}, t_{\rm m}, M|t)$ can be computed again via the path
decomposition method as before by splitting a path into three intervals:
$[0,t_{\rm l}]$, $[t_{\rm l}, t_{\rm m}]$ and $[t_{\rm m}, t]$ in the case when $t_{\rm l}<t_{\rm m}$.
In the opposite case $t_{\rm m}< t_{\rm l}$ we split the path into intervals $[0,t_{\rm m}]$, $[t_{\rm m}, t_{\rm l}]$
and $[t_{\rm l}, t]$ respectively. The propagators of the path in the three intervals
are calculated separately and then multiplied together to obtain the full
joint distribution $P(t_{\rm l}, t_{\rm m}, M|t)$. Finally, using (\ref{marginal_13}), we obtain the
desired result for $P_{13}(t_{\rm l}, t_{\rm m}|t)$. 
Skipping details that can be found in
Appendix \ref{sec:tl:tm}, we present our main result here. We get}
\begin{widetext}
\begin{eqnarray}
P_{13}(t_{\rm l},t_{\rm m}|t)&=& \left[\frac{1}{\pi\, t_{\rm l}^{3/2}\, 
\sqrt{t-t_{\rm l}}}\, \sum_{n=0}^{\infty} \frac{(-1)^n}{(1+\alpha\, n^2)^{3/2}}
\right]\, \theta(t_{\rm l}-t_{\rm m})+ \nonumber \\
&+ & \left[\frac{1}{\pi\, t_{\rm l}^2}\, \sum_{n,m=0}^{\infty}
\frac{(-1)^{(n-1)}\, n^2}{\gamma_{n,m}^{3/2}}\,
\left[1 - \left(1+ 2\,\pi\, \sqrt{\gamma_{n,m}}\, \right)\,
e^{-2\, \pi\, \sqrt{\gamma_{n,m}} }\, \right]\right]\, \theta(t_{\rm m}-t_{\rm l})\, ,
\label{jointP13.1}
\end{eqnarray}
\end{widetext}
with $\gamma_{n,m}= \frac{1}{t_{\rm l}}\,\left[n^2\, (t_{\rm m}-t_{\rm l})+ 
(2m+1)^2\, (t-t_{\rm m})\right]$ and 
$ \alpha= \frac{t_{\rm m}\, (t_{\rm l}-t_{\rm m})}{t_{\rm l}\, (t-t_{\rm l})}$.
\blue{Let us remark that while this joint distribution is explicit and can be plotted
(see below), it is very hard to calculate the moments and correlations between $t_{\rm l}$
and $t_{\rm m}$ from Eq. (\ref{jointP13.1}). At least, we were unable to do it and
this remains an interesting future challenge. Also, even deriving the marginal arcsine laws
for $t_{\rm l}$ and $t_{\rm m}$ from Eq. (\ref{jointP13.1}) is not obvious. However, one can check
numerically that they do reproduce the arcsine laws for the marginals.} 

\blue{The adavantage of the explicit result in Eq. (\ref{jointP13.1}) is that
it can be numerically evaluated and plotted}. Indeed,
we numerically evaluated the double sum in Eq.~(\ref{jointP13.1})
  using the {\tt mpfr} library \cite{mpfr-lib}, for an arbitrary choice
  of $t=8$,
  by summing all terms with $n+m\le 5000$. 
 In Fig.~\ref{fig:tltm}, we plot the resulting
$P_{13}(t_{\rm l},t_{\rm m}|t)$ as a function of $t_{\rm l}$ for three different fixed
 values of $t_{\rm m}$, and the simulation results for comparison. They match
 very well. \blue{For fixed values of $t_{\rm m}$ and $t$, the joint PDF 
$P_{13}(t_{\rm l}, t_{\rm m}|t)$, as a function of $t_{\rm l}$, exhibits a strong peak at 
$t_{\rm l}=0$. This peak corresponds to paths that start at the origin, wander to the positive side,
achieve their maximum at the given $t_{\rm m}$
and do not cross the origin further, so that the last-passage through the origin occurs essentially
at the begining so that $t_{\rm l}=0$.} 

\blue{In Fig.~\ref{fig:tltm}, one also observes a discontinuity of $P_{13}(t_{\rm l}, t_{\rm m}|t)$ as a function 
of $t_{\rm l}$ (for fixed $t_{\rm m}$ and $t$) }. 
While $P_{13}(t_{\rm l}, t_{\rm m}|t)$ vanishes very fast as $t_{\rm l}\to t_{\rm m}$ from below,
 it approaches a nonzero constant when $t_{\rm l}\to t_{\rm m}$ from above.
 This can be understood as follows: For a given time $t_{\rm m}$ of maximum,
 the only possibility that the last passage at zero  occurs slightly
 before the maximum is that the trajectory crosses zero in positive direction,
 reaches the maximum quickly, i.e., not far from position zero, and then stays
 for the rest of the trajectory in the narrow region between the
 small maximum and zero.  This is very unlikely and becomes
 even more unlikely for $t_{\rm l}$ approaching $t_{\rm m}$.
   Therefore $P_{13}(t_{\rm l}, t_{\rm m}|t)\to 0$ vanishes rapidly as $t_{\rm l}$ approaches
$t_{\rm m}$ from below. On the other hand,
 for $t_{\rm l}\to t_{\rm m}$ from above, the last passage is almost sure
 downward to the negative part, thus there are many trajectories
 possible, which leads to a much higher probability, explaining the jump
 in $P_{13}(t_{\rm l}, t_{\rm m}|t)$ as a function of $t_{\rm l}$. This property may hold for other types of
 stochastic processes as well.
 In \blue{Appendix \ref{sec:tl:tm}}, we also analyze the \blue{magnitude} of
 this discontinuity as a function of $t_{\rm m}$.


\section{Conclusions}
\label{conclusion}

We have computed analytically, and verified numerically, the pairwise joint distributions for
three global quantities $t_{\rm o}$ (occupation time), $t_{\rm m}$ (time of the maximum)
and $t_{\rm l}$ (last-passage time) for a one-dimensional Brownian time series of duration $t$. 
The main conclusion is that while the marginal distributions of these three variables
all follow the same arcsine law, their joint distributions display very nontrivial and rich behaviors.
\blue{Of the three pairs, some exhibit positive correlation while others weak anticorrelations.
For example, $t_{\rm o}$ and $t_{\rm m}$ exhibit positive correlations, while
$t_{\rm o}$ and $t_{\rm l}$ display weak higher order anticorrelations. Such behaviors
could not have been guessed without the explicit exact results obtained in this paper. An
interesting open problem is to compute explicitly the correlations between $t_{\rm l}$
and $t_{\rm m}$. Ironically for this pair, it seems easier to obtain the full joint distribution
as presented in the paper rather than the moments of this distribution.}

\blue{To the best of our knowledge, these interesting questions about correlations
between global observables have never been addressed in the literature for any
stochastic process, even though they seem natural enough.
Since Brownian motion is the simplest, and yet the most ubiquitous stochastic process appearing
in many areas from physics to biology, our results
may thus serve as a future benchmark in many applications. In addition, it would be
interesting to address these questions for other processes, e.g.,  a natural one being
the Ornstein-Uhlenbeck (OU) process in one dimension. The OU process corresponds to a overdamped
Brownian particle diffusing in the presence of a harmonic potential. 
Unlike the Brownian motion, the OU process is a stationary process.
For such stationary processes, the occupation
time distribution does not exhibit the arcsine law as in Brownian motion, but rather a large-deviation behavior~\cite{majumdar2002b},
i.e., in the limit $t_{\rm o}$ large, $t$ large but with
their ration $t_{\rm o}/t$ fixed, one gets
\begin{equation}
p_{\rm o}(t_{\rm o}|t) \sim \exp\left[- t\, \Phi\left(\frac{t_{\rm o}}{t}\right)\right]\, 
\label{OU_to}
\end{equation}
where the rate function $\Phi(z)$ can be computed for the OU process~\cite{MB2002}.
Similarly, the marginal distribution $p_{\rm m}(t_m|t)$ also exhibits quite different behavior
compared to the arcsine laws~\cite{MMS21,MMS22}. It would certainly be interesting to compute
joint distributions of these three observables for the OU process.}


Finally, since joint distributions carry much more information about a system
than the marginal distributions, obtaining them for other
stochastic processes will be of high general interest. In particular this will also allow
for practical applications to investigate hidden properties by
exploiting the knowledge of correlations, e.g., between two quantities where
one is easily accessible, while the other is not.

\begin{acknowledgments}
  SNM acknowledges the Alexander von Humboldt foundation for the
  Gay Lussac-Humboldt
  prize that allowed a visit to the Physics department at
  Oldenburg University, Germany where most of this work was performed.
  The simulations were performed at the
  the HPC cluster ROSA, located at the University of Oldenburg
  (Germany) and
    funded by the DFG through its Major Research Instrumentation Program
    (INST 184/225-1 FUGG) and the Ministry of
    Science and Culture (MWK) of the
    Lower Saxony State.
\end{acknowledgments}

\begin{appendix}

\section{Joint distribution of the last-passage time $t_{\rm l}$
and the occupation time $t_{\rm o}$ for a Brownian motion of duration $t$}
\label{sec:tl:to}

\begin{figure}
\centering
\includegraphics[width = 0.95 \linewidth]{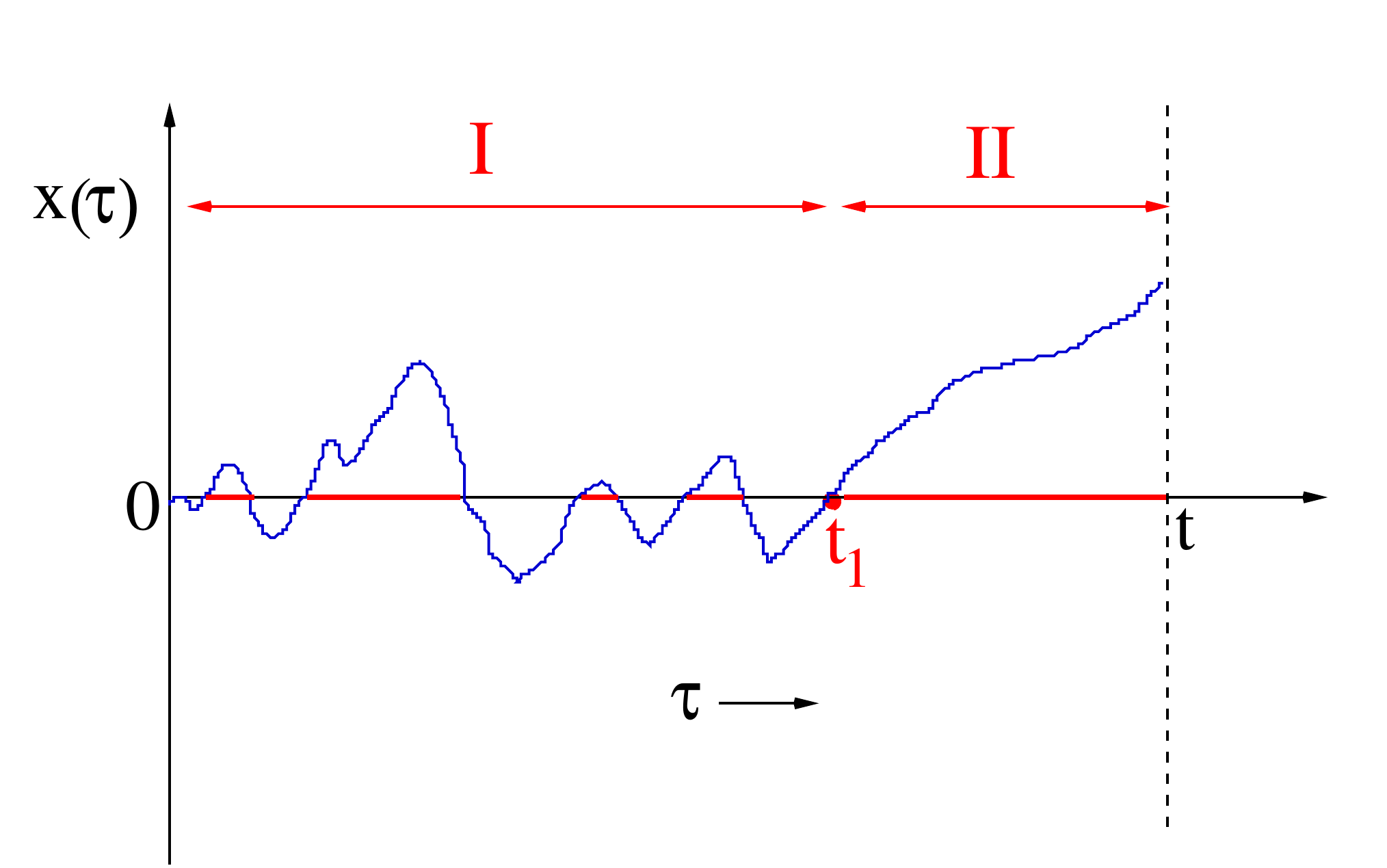}
\caption{\label{Fig1:traj} A typical trajectory of a $1$-d 
Brownian motion $x(\tau)$ of total duration $t$, starting at the 
origin $x(0)=0$ and passing
the origin for the last time before $t$ at $\tau=t_{\rm l}$. Solid red lines show
the intervals during which $x(\tau)$ stays positive. The interval $[0,t_{\rm l}]$
and $[t_{\rm l},t]$ are marked respectively ${\rm I}$ and ${\rm II}$.}
\end{figure}

We consider the Brownian motion $x(\tau)$ of total duration $t$, starting
at $x(0)=0$.
In this section we compute the
joint distribution $P_{12}(t_{\rm l},\,t_{\rm o}|t)$ 
of the last-passage time $t_{\rm l}$ and the occupation time
$t_{\rm o}$ for fixed $t$. 
The process crosses the origin for the last 
time before $t$ at the instant $\tau=t_{\rm l}$ (see Fig. (\ref{Fig1:traj})).

Here we compute exactly the joint distribution $P_{12}(t_{\rm l},t_{\rm o}
|t)$ for a fixed total duration $t$.
To proceed, we consider a typical trajectory of the Brownian motion
as in Fig. (\ref{Fig1:traj}). The idea is to first express the 
joint distribution as
\begin{equation}
P_{12}(t_{\rm l},t_{\rm o}|t)= P_{\rm cond}(t_{\rm o}|t_{\rm l}|t)\, p_1(t_{\rm l}|t)\, ,
\label{cond_def}
\end{equation}
where $P_{\rm cond}(t_{\rm o}|t_{\rm l}|t)$ denotes the conditional
distribution of $t_{\rm o}$ given $t_{\rm l}$ and $t$, while $p_1(t_{\rm l}|t)$ is the
marginal distribution of $t_{\rm l}$ following the arcsine law.
To compute the conditional probability $P_{\rm cond}(t_{\rm o}|t_{\rm l}|t)$,
given $t_{\rm l}$ and $t$, we split the total duration $t$ into two intervals (see Fig.
(\ref{Fig1:traj})): (${\rm I}$) $0\le \tau \le t_{\rm l}$ and (${\rm II}$) 
$t_{\rm l}\le \tau\le t$.
Consequently, $t_{\rm o}$ can also be split into two parts
\begin{equation}
t_{\rm o}= t_{\rm o}^{\rm I} + t_{\rm o}^{\rm II}\, ,
\label{occ_split}
\end{equation}
where the two random variables $t_{\rm o}^{\rm I}$ and $t_{\rm o}^{\rm II}$ are independent
of each other, for a given fixed $t_{\rm l}$. The idea then would be to compute
the probability
distribution function (PDF) of each of them separately and then 
convolute them to compute the full conditional probability.
Below we compute the PDF's of $t_{\rm o}^{\rm I}$ and $t_{\rm o}^{\rm II}$
separately.

\vspace{0.3cm}

\noindent {\bf Interval ${\rm I}$} ($0\le \tau\le t_{\rm l}$). In this interval
the Brownian motion starts
at the origin at $\tau=0$ and ends at the origin at $\tau=t_{\rm l}$. Thus we have
a Brownian bridge of duration $t_{\rm l}$ and we want to compute the PDF
of $t_{\rm o}^{\rm I}= \int_0^{t_{\rm l}} \theta(x(\tau))\, d\tau$.
It is well known that for a bridge of duration $t_{\rm l}$,
the PDF of the occupation time $t_{\rm o}^{\rm I}$ is flat~\cite{Feller}, i.e.,
\begin{equation}
p_{\rm I}(t_{\rm o}^{\rm I}|t_{\rm l})= \frac{1}{t_{\rm l}}\, \theta\left(t_{\rm l}- 
t_{\rm o}^{\rm I}\right)\, .
\label{pleft.1}
\end{equation}
The theta function gurantees that $t_{\rm o}^I \le t_{\rm l}$.
One can also derive this result using the Feynman-Kac equation (by
easily adapting the method described in e.g. Ref.~\cite{BF2005} for
a free Brownian motion). For convenience, we add a 
short derivation of this result later
in the last section of this Supplemental Material.

\vspace{0.3cm}

\noindent {\bf Interval ${\rm II}$} ($0\le \tau\le t_{\rm l}$). 
We first note that when the
process crosses zero for the last time at $\tau=t_{\rm l}$ (see Fig. (\ref{Fig1:traj})),
it may do so either from above or below with equal probability $1/2$.
If it crosses zero from above, then for the interval ${\rm II}$, 
i.e., for $\tau\in [t_{\rm l},t]$,
the process stays below zero indicating $t_{\rm o}^{\rm II}=0$. 
In contrast, if it crosses
from zero from below at $\tau=t_{\rm l}$ for the last time, it stays above zero in
interval ${\rm II}$, indicating $t_{\rm o}^{\rm II}=(t-t_{\rm l})$. 
Since the two events occur
with equal probability $1/2$, the distribution of the occupation time for
the interval ${\rm II}$ is simply given by
\begin{equation}
p_{\rm II}(t_{\rm o}^{\rm II}|t_{\rm l})= \frac{1}{2}\, \delta\left(t_{\rm o}^{\rm II}\right)
+\frac{1}{2}\, \delta\left(t_{\rm o}^{\rm II}- (t-t_{\rm l})\right)\, .
\label{pright.1}
\end{equation}

Since the total occupation time $t_{\rm o}=t_{\rm o}^{\rm I}+t_{\rm o}^{\rm II}$ in Eq. (\ref{occ_split}) is the sum of these two
independent variables $t_{\rm o}^{\rm I}$ and $t_{\rm o}^{\rm II}$ (for fixed given $t_{\rm l}$), 
its conditional
distribution is given by the convolution
\begin{equation}
P_{\rm cond}(t_{\rm o}|t_{\rm l}|t)=  \int_0^t dt_{\rm o}^{\rm I}\, p_{\rm I}(t_{\rm o}^{\rm I})\,
p_{\rm II}(t_{\rm o}-t_{\rm o}^{\rm I})\, .
\label{conv.1}
\end{equation}
Substituting the results from Eqs. (\ref{pleft.1}) and (\ref{pright.1}) in 
Eq. (\ref{conv.1}) gives (after carefully taking into acount when the delta functions
contribute to the integral)
\begin{multline}
  P_{\rm cond}(t_{\rm o}|t_{\rm l}|t)= \\
  \left[ \frac{1}{2\,t_{\rm l}}\, \theta(t_{\rm l}-t_{\rm o})
+\frac{1}{2\,t_{\rm l}}\,\theta(t_{\rm o}-(t-t_{\rm l}))\right]\, \theta(t-t_{\rm o})\, .
\label{pcond.1}
\end{multline}

One can easily verify the normalization
\begin{equation} 
\int_0^t P_{\rm cond}(t_{\rm o}|t_{\rm l}|t)\, dt_{\rm o}=1 \, .
\label{norm.1}
\end{equation}

We now substitute the result from Eq. (\ref{pcond.1}) into Eq. (\ref{cond_def})
and use the arcsine result for the marginal distribution of $t_{\rm l}$.
This gives our final result
for the joint distribution of $t_{\rm l}$ and $t_{\rm o}$
\begin{multline}
  P_{12}(t_{\rm l},t_{\rm o}|t)= \\
  \frac{
\left[ \theta (t_{\rm l}-t_{\rm o})+ \theta(t_{\rm o}-(t-t_{\rm l}))\right]\, \theta(t-t_{\rm o})\, \theta(t-t_{\rm l})}{2\, \pi\, t_{\rm l}^{3/2}\, \sqrt{t-t_{\rm l}}}\, .
\label{jpdf_tlto.1}
\end{multline}
It is convenient to express this joint PDF in the scaling form
\begin{equation}
P_{12}(t_{\rm l},t_{\rm o}|t)= \frac{1}{t^2}\, {\cal P}_{12}\left(\frac{t_{\rm l}}{t}=x,\, 
\frac{t_{\rm o}}{t}=y\right)\, ,
\label{jpdf_tlto_scaled}
\end{equation}
where the scaling function ${\cal P}_{12}(x,y)$ for $0\le x\le 1$ and
$0\le y\le 1$ is given by \eqref{tlto_scaled.1}.
One can check that the scaling function is normalized to unity
\begin{equation}
\int_0^1 \int_0^1 dx\, dy\,{\cal P}_{12}(x,y) =1\, .
\label{scaling_norm.1}
\end{equation}
Furthermore, integrating over $y$ only or $x$ only, one recovers the
marginal arcsine-law distributions,
respectively. Finally, we note that as in the case of marginal distributions,
the joint distribution of $t_{\rm l}$ and $t_{\rm o}$ is also independent of
the diffusion constant $D$ of the Brownian motion.

To visualize this scaled joint PDF (\ref{tlto_scaled.1}) in the $(x,y)$ plane 
(actually in the
square $[0,1]\times [0,1]$), it is useful to fix $y$
and observe the scaled joint PDF as a function of $x$ (along a horizontal slice
in the $(x,y)$ plane). There are two situations $0\le y\le 1/2$ and 
and $1/2<y\le 1$ that we consider separately.

\vspace{0.3cm}

\noindent Case 1: $0\le y\le 1/2$. In this case, since $y\le 1/2$, it is clear
that $y\le 1-y$. We then have three regions as a function of $x$, for fixed 
$0\le y\le 1/2$, where
the scaled joint PDF in Eq. (\ref{tlto_scaled.1}) takes the following form

\change{
  \begin{equation}
{\cal P}_{12}(x,y)= \begin{cases}
& 0\, , \quad\quad\quad\quad\quad\,\,\,\,\,\, {\rm for}\quad 
0\le x < y\, ,  \\
\\
& \frac{1}{2\, \pi\, x^{3/2}\, \sqrt{1-x}}\,\, , \quad\, 
{\rm for} \quad y<x<1-y \,  \\
\\
& \frac{1}{ \pi\, x^{3/2}\, \sqrt{1-x}}\, , \quad\quad {\rm for} 
\quad  1-y<x\le 1 \, .
\end{cases}
\label{P12.1}
\end{equation}}
Thus the scaled joint PDF, as a function of $x$ for fixed $0\le y\le 1/2$, undergoes
two discontinuous jumps as a function of $x$, respectively at $x=y$ and
$x=1-y$.

\vspace{0.3cm}

\noindent Case 2: $1/2\le y\le 1$. In this case, the scaling function in 
Eq. (\ref{tlto_scaled.1}) reads
\begin{equation}
{\cal P}_{12}(x,y)= \begin{cases}
& 0\, , \quad\quad\quad\quad\quad\,\,\,\,\,\, 
{\rm for}\quad 0\le x < 1-y\, ,  \\
\\
& \frac{1}{2\, \pi\, x^{3/2}\, \sqrt{1-x}}\,\, , \quad\, 
{\rm for} \quad 1-y<x<y \,  \\
\\
& \frac{1}{ \pi\, x^{3/2}\, \sqrt{1-x}}\, , \quad\quad {\rm for} 
\quad  y<x\le 1 \, .
\end{cases}
\label{P12.2}
\end{equation}
Here again, for fixed $y$, as a function of $x$, the scaling function
undergoes discontinuous jumps at $x=1-y$ and $x=y$. 

Combining these two behaviors for $y\le 1/2$ and $y\ge 1/2$,
we obtain $4$ distinct triangular regimes in the $2$-d $(x-y)$ square of unit length, that
are separated by the two diagonal lines $y=x$ and $y=1-x$.
In the triangle $(x<y<1-x)\times (x<1/2)$, we have ${\cal P}_{12}(x,y)=0$.
In the triangles $(y<x<1-y)\times (y<1/2)$
and $(1-y<x<y)\times (y>1/2)$, we have ${\cal P}_{12}(x,y)= 1/[2\,\pi\, x^{3/2}\,
\sqrt{1-x}]$. Finally, in the triangle $(1-x<y<x)\times (x>1/2)$, we have
${\cal P}_{12}(x,y)= 1/[\pi\, x^{3/2}\,
\sqrt{1-x}]$.

\vspace {0.3cm}

\noindent {\bf Correlation between $t_{\rm l}$ and $t_{\rm o}$.} Given the joint distribution
function in Eq. (\ref{tlto_scaled.1}) one can compute the joint moments of the form
\begin{equation}
\langle t_{\rm l}^m\, t_{\rm o}^n\rangle= t^2\, \langle x^n\, y^n\rangle= t^2\, \int_0^1 dx\int_0^1 dy 
\, x^m\, y^n \, {\cal P}_{12}(x,y)\, 
\label{jmom.1}
\end{equation}
Interestingly, even though the joint distribution (\ref{tlto_scaled.1}) does not factorise,
the explicit calculation of the covariance shows that it vanishes
\begin{equation}
\langle x\, y\rangle- \langle x\rangle\, \langle y\rangle=0\, .
\label{cov11.1}
\end{equation}
It also turns out that  
\begin{equation}
\langle x^2\, y\rangle-\langle x^2\rangle\, \langle y\rangle=0\ ,
\label{cov21.1}
\end{equation}
which raises the possibility that maybe there is a hidden factorisation
of ${\cal P}_{12}(x,y)$ into a product of a function of $x$ and 
a function of $y$.
To test this, we computed the correlator
\begin{equation}
\langle x\, y^2\rangle-\langle x\rangle\, \langle y^2\rangle=-\frac{1}{48}\, , 
\label{cov12.1}
\end{equation}
which clearly is nonzero, thus ruling out the factorisation. 
We further computed
\begin{equation}
\langle x^2\, y^2\rangle-\langle x^2\rangle\, \langle y^2\rangle= -\frac{7}{384}\, ,
\label{cov22.1}
\end{equation}
exhibiting again a nonzero correlation.
Thus clearly the correlations become nonzero only at higher orders.
Moreover, the negative sign indicates
that $t_{\rm l}$ and $t_{\rm o}$ are weakly anti-correlated.

We have 
verified these analytical predictions numerically to
very high precision in direct Monte Carlo data.
We have performed simulations consisting
of $10^8$ independent simulations for $n_t=1000$ and $n_t=10000$ steps,
respectively with step size
$\Delta t=1$.
We obtained
for (\ref{cov12.1}) a value of $-0.0209(1)$,
where the error bar reflects the difference between the two results for the
two values of $n_s$. This compares well with $-1/48 \approx -0.02083$.
For (\ref{cov22.1}) we obtained 0.01820(3), compared to
$-7/384 \approx -0.01823$.

\section{Joint distribution of the occupation time $t_{\rm o}$ 
  and the time $t_{\rm m}$ of the maximum for a Brownian motion of duration $t$}
\label{sec:to:tm}

In this section, our goal is to compute the joint distribution
$P_{23}(t_{\rm o},t_{\rm m}|t)$ of the random variables $t_{\rm o}$ and $t_{\rm m}$, given
the fixed total duration $t$. Here our result is less explicit.
Let us summarize our main result. We first define the 
following triple Laplace transform of the joint
distribution $P_{23}(t_{\rm o}, \, t_{\rm m}|t)$ as

\begin{multline}
\int_0^{\infty} dt\, e^{-s\, t}\,\int_0^{t} dt_{\rm o}\, 
e^{-p\, t_{\rm o}}\,
\int_0^{t} dt_{\rm m}\, e^{- \lambda\, t_{\rm m}} 
P_{23}(t_{\rm o},\, t_{\rm m}|t)= \\
\int_0^{\infty} dt\, e^{-s\,t}\,
\Big\langle e^{-p\, t_{\rm o}- \lambda\, t_{\rm m}}\Big \rangle\, .
\label{triple_lt_def}
\end{multline}
For this quantity, we find the following exact result
\begin{widetext}
\begin{multline}
\int_0^{\infty} dt\, e^{-s\,t}\,
\Big\langle e^{-p\, t_{\rm o}- \lambda\, t_{\rm m}  }\Big \rangle=\\
\frac{1}{(s+p)}\, \int_0^{\infty} dx\,
\frac{ \left[\sinh(x)+ \sqrt{\frac{s}{s+p}}\, \cosh(x)+
\frac{p}{s}\, \sqrt{\frac{s}{s+p}}\right]}
{\left[\cosh(x)+ \sqrt{\frac{s}{s+p}}\, \sinh(x)\right]\,
\left[\sqrt{\frac{s+\lambda}{s+p+\lambda}}\,
\sinh\left(\sqrt{\frac{s+p+\lambda}{s+p}}\, x\right)+
\cosh\left(\sqrt{\frac{s+p+\lambda}{s+p}}\, x\right)
\right]}\, .
\label{tlpt.1}
\end{multline}

\end{widetext}

Even though it is hard to invert this triple Laplace tarnsform to
extract the joint distribution $P_{23}(t_{\rm o},t_{\rm m}|t)$ explicitly, the result
in Eq. (\ref{tlpt.1}) gives us access to exact moments and correlation
functions as shown below.
The derivation of this principal result in Eq. (\ref{tlpt.1}) is
somewhat involved. We will just outline below the main steps and
the readers not interested in the derivation may skip this part
and move directly to the explicit results on moments and correlation functions
derived later from this formula.

\begin{figure}
\centering
\includegraphics[width = 0.95 \linewidth]{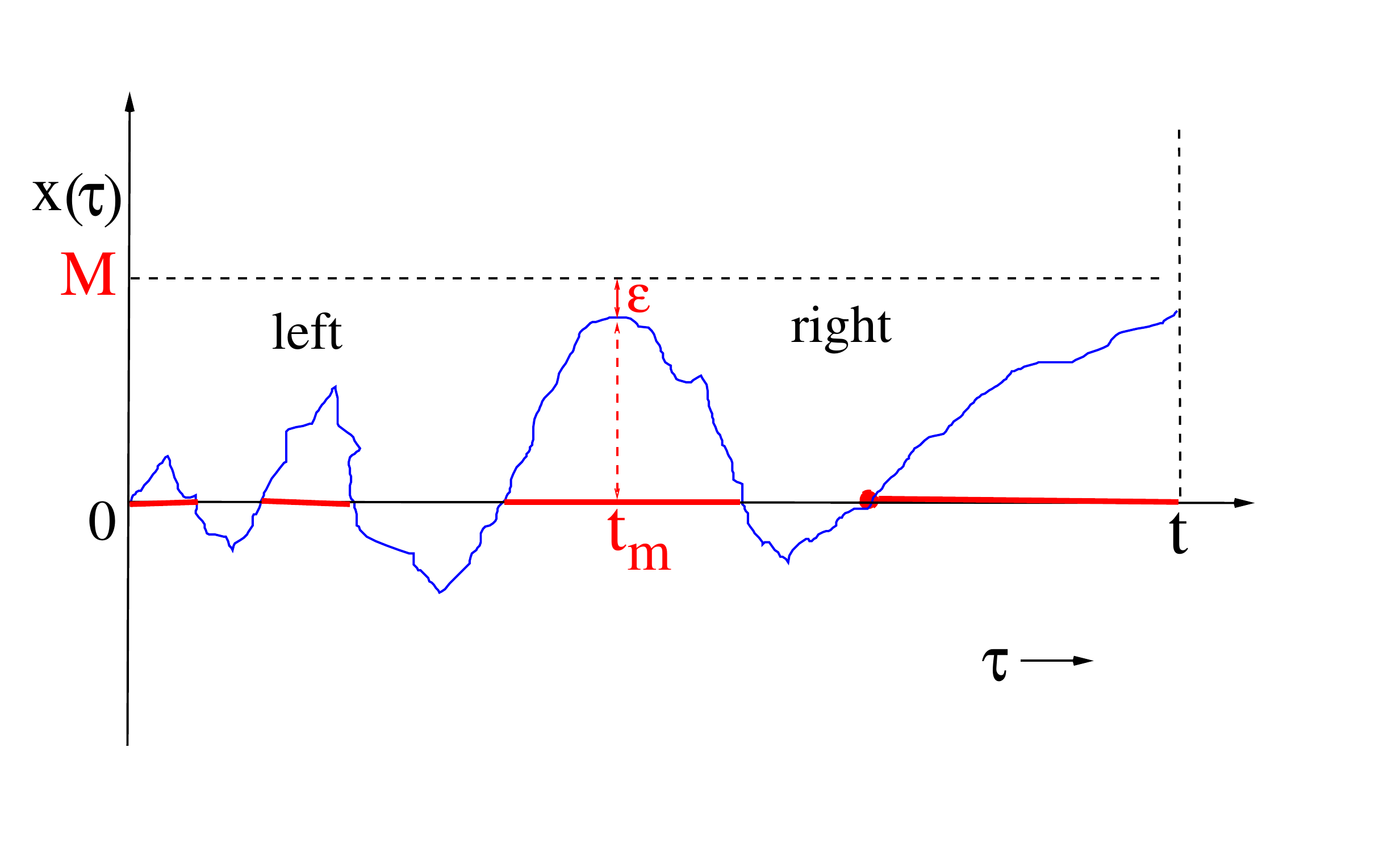}
\caption{\label{Fig23:traj} A typical trajectory of a $1$-d
Brownian motion $x(\tau)$ of total duration $t$, starting at the
origin $x(0)=0$ and achieving it maximum value $M-\epsilon$
at time $t_{\rm m}$.
Solid red lines show
the intervals during which $x(\tau)$ stays positive. The interval $[0,t_{\rm m}]$
and $[t_{\rm m},t]$ are marked respectively ${\rm left}$ and ${\rm right}$.
Both on the left and right regions, the trajectory stays below
the level $M$, since by definition $M$ is the maximum. The cut-off
$\epsilon$ is introduced for the convenience of calculation
and eventually we take the limit $\epsilon\to 0$.}
\end{figure}

To compute the joint distribution $P_{23}(t_{\rm o},t_{\rm m}|t)$, we generalise
a path-decomposition method that was originally used in Ref.~\cite{MRKY08}
to compute the joint distribution of the global maximum $M$ and the time $t_{\rm m}$
at which it occurs for a Brownian motion of duration $t$. 
The idea is
to first compute an extended joint distribution $P(t_{\rm o}, t_{\rm m}, M|t)$
of the three random variables $t_{\rm o}$, $t_{\rm m}$ and the value of the
maximum $M$ and then integrate over $M$ to obtain the joint distribution
of $t_{\rm o}$ and $t_{\rm m}$
\begin{equation}
P_{23}(t_{\rm o},t_{\rm m}|t)= \int_0^{\infty} P(t_0, t_{\rm m}, M|t)\, dM\, .
\label{3rv.1}
\end{equation}

The computation of $P(t_0, t_{\rm m}, M|t)$ can be performed using a path 
decomposition method that exploits the Markov property of the
process and is best understood by following a trajectory as
depicted in Fig. (\ref{Fig23:traj}). We first divide the full
interval $[0,t]$ into two parts: (i) `left' part over $[0,t_{\rm m}]$
and (ii) `right' part over $[t_{\rm m},t]$. In the left part, the trajectory
$x(\tau)$, starting at $0$, has to remain below $M$ till time $t_{\rm m}$
and it arrives at its maximal value $M$ exactly at the instant $t_{\rm m}$.
On the right side, the process, starting at $M$ exactly at $t_{\rm m}$
needs to stay below $M$ till time $t$. Normally, the constraint
of staying below $M$ can be imposed by putting an absorbing boundary
at the level $M$. However, imposing an absorbing boundary at $x=M$
during $[0,t]$
also forbids the process to reach exactly at $M$ at time $t_{\rm m}$.
To satisfy both constraints, i.e., staying below $M$ during the
full interval $[0,t]$ and at the same time arriving at $M$
at the instant $t=t_{\rm m}$, one can use the trick of putting
a small cut-off $\epsilon$ and imposing the condition
that the process reaches $M-\epsilon$ at $t=t_{\rm m}$, while staying
below $M$ through out the interval $[0,t]$. Then there is
no problem in imposing an absorbing boundary condition at $x=M$
during the full interval $[0,t]$.
Finally, we will take the limit $\epsilon\to 0$ in an appropriate way
at the very end of the calculation. This limiting procedure
using an $\epsilon$ cut-off has been successfully used in many examples
of computing observables for constrained Brownian motions~\cite{MC2005,
MRKY08,PCMS13,PCMS15,MMS19,MMS20,MMS21,MMSS21,MMS22}, and we will
henceforth call it the $\epsilon$-path decomposition method. 

In the current problem, in addition to $t_{\rm m}$ and $M$, we need also to 
keep track
of the occupation time $t_{\rm o}$. Let us also divide it into two parts, 
one from the left of $t_{\rm m}$ and one from the right of $t_{\rm m}$, i.e.,
\begin{multline}
t_{\rm o}= \int_0^t \theta\left(x(\tau)\right)\, d\tau=\\ 
\int_0^{t_{\rm m}} \theta\left(x(\tau)\right)\, d\tau
+ \int_{t_{\rm m}}^t \theta\left(x(\tau)\right)\, d\tau= t_{\rm o}^{\rm L}+t_{\rm o}^{\rm R}\, ,
\label{occ_time_decom}
\end{multline}

where the superscripts ${\rm L}$ and ${\rm R}$ refer respectively to the
left and right of $t_{\rm m}$. For the left part, let $G_M
(M-\epsilon, t_{\rm m}, t_{\rm o}^{\rm L}|0,0)$
denote the restricted propagator of the Brownian motion from the position $0$
at time $0$ to position $M-\epsilon$ at time $t_{\rm m}$ and carrying
an occupation time $t_{\rm o}^{\rm L}$, in the presence an absorbing
boundary at $x=M$ (see Fig (\ref{Fig23:traj})). 
Similarly, for the right part, let
$Q_M(M-\epsilon, t-t_{\rm m}, t_{\rm o}^{\rm R})$ denote the probability
that the process, starting at $M-\epsilon$ at time $t_{\rm m}$ will
stay below the level $M$ up to time $t$ and will carry
an occupation time $t_{\rm o}^{R}$ (see Fig. (\ref{Fig23:traj})).
Then using the Markov property of the process, we have
  \begin{multline}
    P(t_{\rm o},t_{\rm m},M|t)= {\cal N}(\epsilon)\times \\
    \int_0^{t_{\rm o}} dt_{\rm o}^{\rm L}\, 
G_M(M-\epsilon, t_{\rm m}, t_{\rm o}^{\rm L}|0,0)\, Q_M(M-\epsilon, t-t_{\rm m}, 
t_{\rm o}-t_{\rm o}^{\rm L})\, ,
\label{3rv.2}
  \end{multline}
where we used the decomposition in Eq. (\ref{occ_time_decom}).
We have introduced a proportionality constant ${\cal N}(\epsilon)$
which will be fixed later from the normalization of the
joint probability density, once we take the $\epsilon\to 0$ limit
at the end of the calculation.
Due to the convolution structure
of the right hand side (rhs) of Eq. (\ref{3rv.2}) with respect to both
$t_{\rm o}$ and $t_{\rm m}$, it is
then natural to define
a Laplace transform with respect to all
three variables $t_{\rm o}$, $t_{\rm m}$ and $t$ 
  \begin{multline}
    \tilde{P}(M, p,\lambda, s)= \\
    \int_0^{\infty} dt\, e^{-s\, t}\, 
\int_0^{t} dt_{\rm o}\, e^{-p\, t_{\rm o}}\, \int_0^t dt_{\rm m}\, e^{-\lambda\, t_{\rm m}}\,
P(t_{\rm o},t_{\rm m},M|t)\, .
\label{3rv_lap.1}
  \end{multline}

Taking this triple Laplace transform in Eq. (\ref{3rv.2}) gives
\begin{equation}
\tilde{P}(M, p,\lambda, s)={\cal N}(\epsilon)\, \tilde{G}_M(M-\epsilon, \lambda+s,p)\, 
\tilde{Q}_M(M-\epsilon, s, p) \, .
\label{3rv_lap.2}
\end{equation}
Here we have defined
\begin{equation}
\tilde{G}_M(x, s, p)= \int_0^{\infty} dT e^{-s\, T}
\int_0^t dt_{\rm o}\, e^{-p\, t_{\rm o}}\, G_{M}(x, T, t_{\rm o}|0,0),
\label{prop_left.1}
\end{equation}
where $G_{M}(x, T, t_{\rm o}|0,0)$ (with $x<M$) is simply the propagator of a
Brownian motion from $(0,0)$ to $(x,T)$ with occupation time $t_{\rm o}$ in $[0,T]$ 
and in the presence of an absorbing boundary at $x=M$.
Similarly
\begin{equation}
\tilde{Q}_M(x_0, s, p)= \int_0^{\infty} dT\, e^{-s\, T}\,
\int_0^t dt_{\rm o}\, e^{-p\, t_{\rm o}}\, Q_M(x_0, T, t_{\rm o})\, ,
\label{prop_right.1}
\end{equation}
where $Q_M(x_0, t_{\rm o}, T)$ is the probability that a Brownian path, starting at
$x_0<M$ at time $0$, stays below $M$ up to time $T$ and carries 
an occupation time $t_{\rm o}$ in $[0,T]$.  
 
Thus we need two auxiliary ingredients, namely the two restricted
propagators $\tilde{G}_M(x, s, p)$ and $\tilde{Q}_M(x_0, s, p)$, in
order to compute the rhs of Eq. (\ref{3rv_lap.2}). Fortunately,
these two restricted propagators (rather their Laplace transforms)
can be computed using the Feynman-Kac formalism. For example, consider
first $G_{M}(x, T, t_{\rm o}|0,0)$ (with $x<M$)
and define its Laplace transform
with respect to $t_{\rm o}$ only, i.e.,
\begin{equation}
\tilde{g}_M(x,T, p)= \int_0^T dt_{\rm o}\,  e^{-p\, t_{\rm o}}\, G_{M}(x, T, t_{\rm o}|0,0)\, .
\label{single_lap.1}
\end{equation} 
From the definition $t_{\rm o}= \int_0^T  
\theta\left(x(\tau)\right)\, d\tau$, we thus have a functional of Brownian
motion and consequently $\tilde{g}_M(x,T, p)$ satisfies
the Feynman-Kac equation
\begin{equation}
\partial_T \tilde{g}_M(x,T, p)= D\, \partial_x^2 \tilde{g}_M(x,T, p)
-p\, \theta(x)\, \tilde{g}_M(x,T, p)\, ,
\label{FK.1}
\end{equation}
valid for $x\le M$ with the absorbing boundary condition
$\tilde{g}_M(x=M,T, p)=0$. The initial condition
reads $\tilde{g}_M(x,T=0, p)= \delta(x)$,
since the process starts at the origin. Next we take a Laplace transform with 
respect to $T$ and define  
\begin{equation}
\tilde{G}_M(x,s,p)= \int_0^{\infty} dT\, e^{-s\, T}\, \tilde{g}_M(x,T, p)\, .
\label{def_g.1}
\end{equation}
Taking Laplace transform of Eq. (\ref{FK.1}) with respect to $T$
and using the initial condition gives us an ordinary
second order differential equation for $\tilde{G}_M(x,s,p)$
\begin{equation}
D\, \frac{d^2 \tilde{G}_M(x,s,p)}{dx^2} -\left[p\, 
\theta(x)+s\right]\, \tilde{G}_M(x,s,p)= -\delta(x)\, 
\quad\, 
\label{FK.2}
\end{equation}
for $ x\le M$ ,
with the boundary conditions $\tilde{G}_M(x=M,s, p)=0$ and 
$\tilde{G}_M(x\to -\infty,s, p)=0$. The differential equation
can be solved exactly (in two regions $0\le x \le M$ and $x\le 0$
and then by matching the two solutions at $x=0$). Omitting
details, we get
\begin{multline}
  \tilde{G}_M(x,s,p)=\\
  \frac{\sinh\left(\sqrt{\frac{s+p}{D}}\, (M-x)\right)}
{\sqrt{D}\, \left[ \sqrt{s}\, \sinh\left(\sqrt{\frac{s+p}{D}}\, M\right)
 + \sqrt{s+p}\, \cosh \left(\sqrt{\frac{s+p}{D}}\, M\right)\right]}\, .
\label{FK_sol_left.1}
\end{multline}
Now setting $x=M-\epsilon$ and taking $\epsilon\to 0$ limit, we obtain to
leading order in $\epsilon$
\begin{multline}
\tilde{G}_M(M-\epsilon,s,p) = \\
\frac{\sqrt{s+p}}{D\, \left[ \sqrt{s}\, 
\sinh\left(\sqrt{\frac{s+p}{D}}\, M\right)
 + \sqrt{s+p}\, \cosh \left(\sqrt{\frac{s+p}{D}}\, M\right)\right]}\, 
\epsilon
\label{lin_eps.1}
\end{multline}
plus Terms of order $O(\epsilon^2)$.
Finally, replacing $s$ by $\lambda+s$ as required in Eq. (\ref{3rv_lap.2}),
we get for the left half (of $t_{\rm m}$) the propa\-gator (up to order $\epsilon$)
\begin{widetext}
\begin{equation}
\tilde{G}_M(M-\epsilon,\lambda+s,p)\approx
\frac{\sqrt{s+\lambda+p}}{D\, \left[ \sqrt{\lambda+s}\, 
\sinh\left(\sqrt{\frac{\lambda+s+p}{D}}\, M\right)
 + \sqrt{\lambda+s+p}\, \cosh \left(\sqrt{\frac{\lambda+s+p}{D}}\, 
M\right) \right]}\, 
\epsilon \, .
\label{left_sol.1}
\end{equation}
\end{widetext}

We now turn to the right part of $t_{\rm m}$. 
Here we again define first the Laplace transform 
of $Q_M(x_0, t_{\rm o}, T)$ with respect to $t_{\rm o}$ only
\begin{equation}
\tilde{q}(x_0,T,p)= \int_0^{T} dt_{\rm o}\, e^{-p\, t_{\rm o}}\, Q_M(x_0, T, t_{\rm o})\, .
\label{BFK.1}
\end{equation}
Then, $\tilde{q}(x_0,T,p)$ satisfies the backward Feynman-Kac equation
(where one treats the starting position $x_0$ as the variable~\cite{BF2005})
\begin{equation}
\partial_T \tilde{q}_M(x_0,T,p)= D\, \partial_{x_0}^2 \tilde{q}_M(x_0,T, p)
-p\, \theta(x_0)\, \tilde{q}_M(x_0,T, p)\, ,
\label{BFK.2}
\end{equation}
valid for $x_0\le M$ with an absorbing boundary condition 
$\tilde{q}_M(x_0=M,T,p)=0$. The initial condition is
$\tilde{q}_M(x_0,T=0,p)=1$ for all $x_0<M$. Taking again
Laplace transform with respect to $T$ reduces the 
partial differential equation to an ordinary second order 
differential equation for $\tilde{Q}_M(x_0,s,p)= \int_0^{\infty} dT\,
e^{-s\, T}\, \tilde{q}_M(x_0,T,p)$, leading to
\begin{equation}
D\, \frac{d^2 \tilde{Q}_M(x_0,s,p)}{dx_0^2} -\left[p\,
\theta(x_0)+s\right]\, \tilde{Q}_M(x_0,s,p)= -1
\label{BFK.3}
\end{equation}
for  $x\le M$. Once again, one can solve this
differential equation explicitly, with a
long and ugly (but explicit) solution. However, if one now sets 
$x_0=M-\epsilon$ as required in Eq. (\ref{3rv_lap.2}) and expands for small
$\epsilon$, one gets a relatively simpler expression for the leading
order in $\epsilon$
\begin{widetext}
\begin{equation}
\tilde{Q}_M(m-\epsilon,s,p)\approx
\frac{\epsilon}{\sqrt{D\,(s+p)}}\, 
\frac{ \left[ \sinh\left(\sqrt{\frac{s+p}{D}}\, M\right)+
\sqrt{\frac{s}{s+p}}\, \cosh \left(\sqrt{\frac{s+p}{D}}\, M\right)+ 
\frac{p}{\sqrt{s(s+p}}
\right] }{ \left[\cosh \left(\sqrt{\frac{s+p}{D}}\, M\right)+ 
\sqrt{\frac{s}{s+p}}\, \sinh\left(\sqrt{\frac{s+p}{D}}\, M\right)\right]}\, .
\label{right_sol.1}
\end{equation}

Finally, substituting the results from Eq. (\ref{left_sol.1})
and (\ref{right_sol.1}) into Eq. (\ref{3rv_lap.2}) and taking
the limit $\epsilon\to 0$ gives 
\begin{equation}
\tilde{P}(M, p,\lambda, s)=  \frac{ A_1\, \left[\sinh\left(\sqrt{\frac{s+p}{D}}\, M\right)+
\sqrt{\frac{s}{s+p}}\, \cosh \left(\sqrt{\frac{s+p}{D}}\, M\right)+
\frac{p}{\sqrt{s\,(s+p)}}
\right] }{ \left[ \sqrt{\frac{\lambda+s}{\lambda+s+p}}\, 
\sinh\left(\sqrt{\frac{\lambda+s+p}{D}}\, M\right)+ 
\cosh \left(\sqrt{\frac{\lambda+s+p}{D}}\,
M\right)\right]\,\left[\cosh \left(\sqrt{\frac{s+p}{D}}\, M\right)+
\sqrt{\frac{s}{s+p}}\, \sinh\left(\sqrt{\frac{s+p}{D}}\, M\right)\right]}\, ,
\label{final_sol.1}
\end{equation}
\end{widetext}
where the constant $A_1$ is given by
\begin{equation}
A_1=\frac{1}{D\, \sqrt{D\, (s+p)}}\, 
\left[\lim_{\epsilon\to 0} {\cal N}(\epsilon)\,
\epsilon^2\right]\, .
\label{def_A}
\end{equation}
One can check a posteriori that in order that $P(M, t_{\rm o},t_{\rm m}|t)$ is normalized to unity when intergrated over $M$, 
$t_{\rm m}$ and 
$t_{\rm o}$, we must have the
identity
\begin{equation}
\lim_{\epsilon\to 0} {\cal N}(\epsilon)\,
\epsilon^2 = D\,  \quad\, {\rm implying}\quad A_1= \frac{1}{\sqrt{D\, (s+p)}}\, .
\label{norm_fix.1}
\end{equation}
We use this result in Eq. (\ref{final_sol.1}) and then 
integrate over $M$ as in Eq. (\ref{3rv.1}), with the change of variable
$M \sqrt{(s+p)/D}=x$. This then yields the final result
announced in Eq. (\ref{tlpt.1}), thus completing its derivation.

\vspace{0.4cm}

\noindent{\bf Marginals, moments and correlation functions:}
Our main result in this section, namely Eq. (\ref{tlpt.1}),
is exact but somewhat formal. We now show how to exploit
this exact result (\ref{tlpt.1}) on the triple Laplace transform
to extract as much \blue{information} as possible, which can then be
checked numerically. 

\vspace{0.3cm}

\noindent{\bf Marginals.} As a first check, we consider the marginals.
Setting $p=0$ in Eq. (\ref{tlpt.1}) and carrying out the integral
over $x$ on the rhs of Eq. (\ref{tlpt.1}) gives
\begin{equation}
\int_0^{\infty} dt\, e^{-st}\, \int_0^t dt_{\rm m}\, e^{-\lambda\, t_{\rm m}}\, P(t_{\rm m}|t)
=\frac{1}{\sqrt{s\,(s+\lambda)}}\, ,
\label{marg_tm.1}
\end{equation}
where $P(t_{\rm m}|t)= {\rm Prob.}[t_{\rm m}|t]$ is the marginal distribution
of $t_{\rm m}$ for fixed total duration $t$. To invert the double Laplace
transform in Eq. (\ref{marg_tm.1}), we first write $t_1= t-t_{\rm m}$
and split $t= t_{\rm m}+t_1$. Then Eq. (\ref{marg_tm.1}) reduces to
\begin{equation}
\int_0^{\infty} dt_{\rm m}\, e^{-(s+\lambda)\, t_{\rm m}} \int_0^{\infty} dt_1\, 
e^{-s\, t_1}\, P(t_{\rm m}|t)= \frac{1}{\sqrt{s\,(s+\lambda)}}\, .
\label{marg_tm.2}
\end{equation}
Now, the inverse Laplace transform of $1/\sqrt{s}$ is simply given by
${\cal L}^{-1}_{s\to t_1}= 1/\sqrt{\pi\, t_1}$. Using this
we invert both Laplace transforms separately in Eq. (\ref{marg_tm.2})
we recover the well known arcsine law for the marginal distribution of $t_{\rm m}$
\begin{equation}
\change{P(t_{\rm m}|t)}= \frac{1}{\pi}\, \frac{1}{\sqrt{t_{\rm m}\, (t-t_{\rm m})}}\, 
\quad {\rm with}\quad 0\le t_{\rm m}\le t\, .
\label{marg_tm.3}
\end{equation}
Similarly, setting $\lambda=0$ in Eq. (\ref{tlpt.1}), one can again
carry out the integral over $x$ explicitly on the rhs of 
Eq. (\ref{tlpt.1}), leading to
\begin{equation}
\int_0^{\infty} dt\, e^{-st}\, \int_0^t dt_{\rm o}\, e^{-p\, t_{\rm o}}\, P(t_{\rm o}|t)
=\frac{1}{\sqrt{s\,(s+p)}}\, ,
\label{marg_to.1}
\end{equation}
where $P(t_{\rm o}|t)= {\rm Prob.}[t_{\rm o}|t]$ is the marginal distribution
of $t_{\rm o}$ for fixed $t$.
This double Laplace transform in Eq. (\ref{marg_to.1}) can then be
inverted using the same trick as in Eq. (\ref{marg_tm.2}) (just replacing
$\lambda$ by $p$) and thus one again recovers the arcsine law
for the marginal distribution of $t_{\rm o}$
\begin{equation}
P(t_{\rm o}|t)= \frac{1}{\pi}\, \frac{1}{\sqrt{t_{\rm o}\, (t-t_{\rm o})}}\, 
\quad {\rm with}\quad 0\le t_{\rm o}\le t\, .
\label{marg_to.2}
\end{equation}
The fact that we recover the correct normalized marginals also
proves, a posteriori, the normalization condition in Eq. (\ref{norm_fix.1}).

\vspace{0.3cm}

\noindent{\bf Moments and correlation functions.}
For general $p$ and $\lambda$,
it is very hard to invert the triple Laplace transform in
Eq. (\ref{tlpt.1}) explicitly. 
However, we were able to
extract various cross-moments by expanding both sides of Eq. (\ref{tlpt.1})
in powers of $p$ and $\lambda$ and matching their coefficients.
These coefficients are still functions of $s$ and we were able
to invert these Laplace transforms with respect to $s$ explicitly. 
This can be systematically done using the Mathematica.
This allowed us to compute various covariance functions explicitly. 
We quote here some examples.
\begin{eqnarray}
\langle t_{\rm o}\, t_{\rm m}\rangle-
\langle t_{\rm o}\rangle\,\langle t_{\rm m}\rangle &= & \frac{11}{144}\, t^2\, ,
\label{cov11.2} \\
\langle t_{\rm o}\, t_{\rm m}^2\rangle-\langle t_{\rm o}\rangle\, 
\langle t_{\rm m}^2\rangle &=& \frac{43}{648}\, t^3\, ,
\label{cov12.2} \\
\langle t_{\rm o}^2\, t_{\rm m}\rangle-
\langle t_{\rm o}^2\rangle\, \langle t_{\rm m}\rangle &=& 
\frac{509}{7200}\, t^3 \, ,
\label{cov21.2} \\
\langle t_{\rm o}^2\, t_{\rm m}^2\rangle-
\langle t_{\rm o}^2\rangle\, \langle t_{\rm m}^2\rangle 
&= & \frac{25997}{414720}\, t^4 \, .
\label{cov22.2}
\end{eqnarray}
Thus the observables $t_{\rm o}$ and $t_{\rm m}$ are thus slightly positively correlated.
We have verified these analytical predictions numerically,
from the direct Monte Carlo data. We have performed simulations consisting
of $10^8$ independent simulations for $n_t=1000$ and $n_t=10000$ steps,
respectively with step size
$\Delta t=1$. We obtained at $t=1$, 
for (\ref{cov11.2}) a value of $0.0764(1)$,
where the error bar reflects the difference between the two results for the
two values of $n_s$. This compares well with $11/144 \approx 0.07639$.
For (\ref{cov12.2}) we obtained 0.0664(1), compared to
$43/648 \approx 0.06636$. 
For (\ref{cov21.2}) we obtained 0.0707(1), compared to $509/7200 \approx 0.07069$.
For (\ref{cov22.2}) we obtained 0.0627(1),
compared to $25997/414720 \approx 0.06269$.

\vspace{0.3cm}

\section{Joint distribution of the last-passage time $t_{\rm l}$
and the time $t_{\rm m}$ of the maximum for a Brownian motion of duration $t$}
\label{sec:tl:tm}

In this section we compute the joint distribution $P_{13}(t_{\rm l},t_m|t)$
of the two random variables $t_{\rm l}$ and $t_{\rm m}$, given
the fixed total duration $t$. To compute this joint distribution
via the $\epsilon$-path decomposition method as in the previous section,
it is convenient to introduce the enlarged joint distribution
$P(t_{\rm l},t_{\rm m},M|t)$ of $t_{\rm l}$, $t_{\rm m}$ and the value $M$ of the
global maximum in $[0,t]$. Once we obtain this, one can integrate
over $M$ to obtain $P_{13}(t_{\rm l},t_{\rm m}|t)$, i.e.,
\begin{equation}
P_{13}(t_{\rm l},t_{\rm m}|t)= \int_0^{\infty} P(t_{\rm l}, t_{\rm m}, M|t)\, dM\, .
\label{tltm_3rv.1}
\end{equation}

To compute $P(t_{\rm l},t_{\rm m},M|t)$, we proceed as follows.
It turns out that the results are quite different depending
on whether $t_{\rm l}>t_{\rm m}$ or $t_{\rm l}<t_{\rm m}$. We start with the case
$t_{\rm l}>t_{\rm m}$ and then proceed to $t_{\rm l}<t_{\rm m}$.

\vspace{0.3cm}

\begin{figure}
\centering
\includegraphics[width = 0.95 \linewidth]{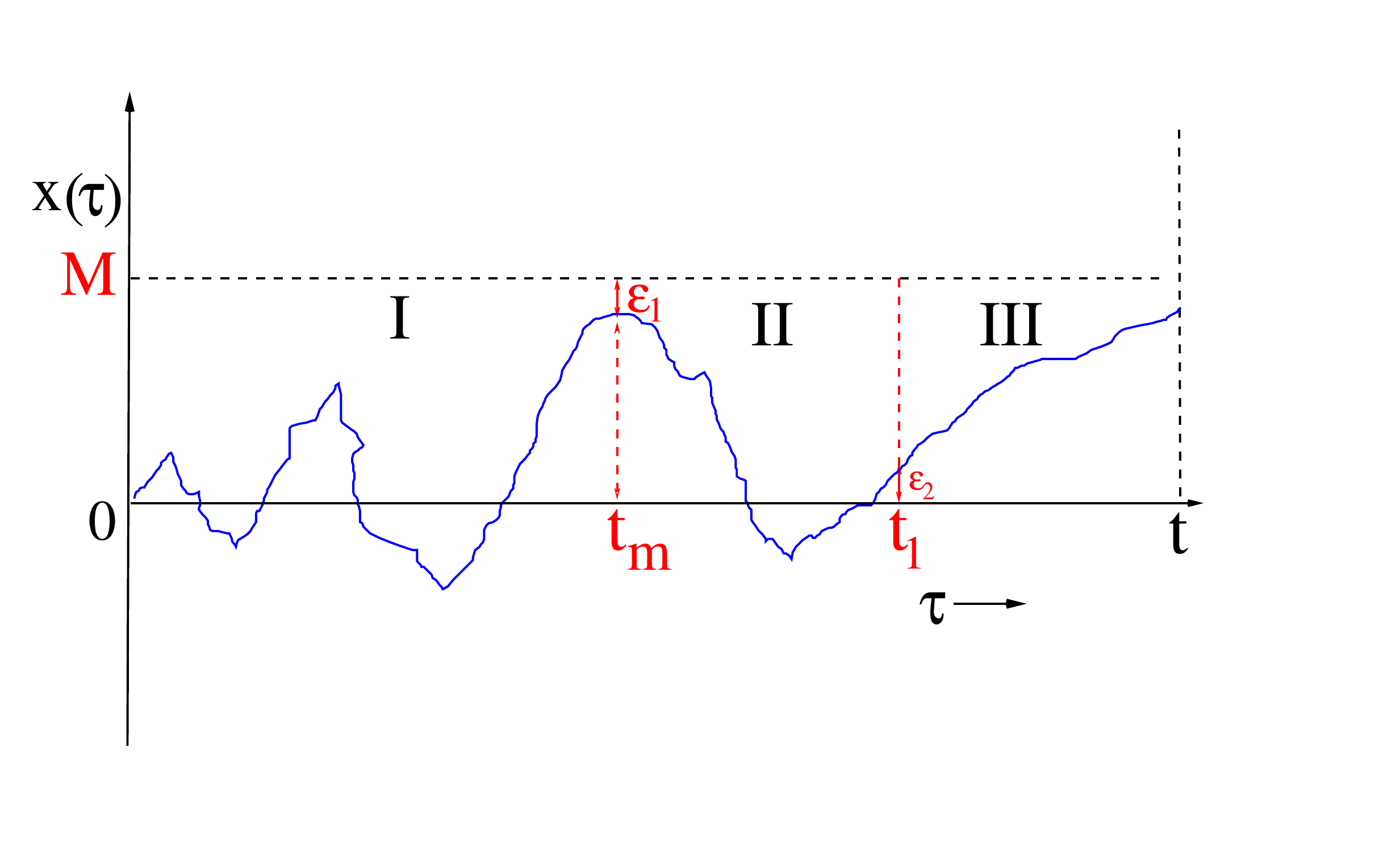}
\caption{\label{Fig13:traj1} A typical trajectory of a $1$-d
Brownian motion $x(\tau)$ of total duration $t$, starting at the
origin $x(0)=0$. The trajectory
arrives at its maximum value $M-\epsilon_1$ at time $t_{\rm m}$
and arrives at $\epsilon_2$ at time $t_{\rm l}>t_{\rm m}$ after which
it does not cross the origin.  We divide
the total duration $[0,t]$ into three intervals:
(${\rm I}$) the interval $[0,t_{\rm m}]$ where the process, starting at the origin
and staying below the level $M$ arrives at $M-\epsilon_1$
at time $t_{\rm m}$ (${\rm II}$) starting at $M-\epsilon_1$ at time $t_{\rm m}$ the
process arrives at $\epsilon_2>0$ at time $t_{\rm l}$, while 
staying below $M$ and (${\rm III}$) starting
at $\epsilon_2$ the process stays below $M$ and does not cross the origin
during $[t_{\rm l}, t]$. Eventually, we take the limits $\epsilon_1\to 0$
and $\epsilon_2\to 0$ at the end of the calculation.}
\end{figure}

\noindent{\bf The case $t_{\rm l}>t_{\rm m}$.} To compute $P(t_{\rm l},t_{\rm m},M|t)$ in this case,
we again investigate a typical trajectory as shown in Fig. (\ref{Fig13:traj1}). 
Let $M$ be the maximum of the Brownian motion of duration $t$ that starts
at the origin $x(0)=0$. Let $t_{\rm m}$ and $t_{\rm l}$ denote respectively
the time at which the maximum occurs and the last time the
processes crosses $0$ before $t$. 
For $M$ to be the maximum, the process has to stay below $M$ throughout
which is implemented by an absorbing boundary condition for
the trajectory at $x=M$. However, this forbids the process
to arrive at $M$ at time $t_{\rm m}$. Hence, as in the
previous ssection, we put a cut-off $\epsilon_1$ at $t=t_{\rm m}$
and impose that the process arrives at 
$M-\epsilon_1$ (with $\epsilon_1\ge 0$) at time $t_{\rm m}$.
Similarly, to impose the constraint that the process stays
above or below the origin during $[t_{\rm l},t_{\rm m}]$ we need to
put an absorbing boundary condition at $x=0$ which will again
forbid the process to arrive at $0$ exactly at $t_{\rm l}$.
Hence, we need to introduce a second cut-off $\epsilon_2$ at $t=t_{\rm l}$,
i.e., impose that the process arrives at $\epsilon_2$ at $t=t_{\rm l}$
and then does not cross zero during $[t_{\rm l},t]$. Eventually, we will
take the limits $\epsilon_1\to 0$ and $\epsilon_2\to 0$
in appropriate ways as in the previous section. 

To proceed with the computation we   
then divide the full time interval $[0,t]$ into three subintervals (see
Fig. (\ref{Fig13:traj1})):
(${\rm I}$) the interval $[0,t_{\rm m}]$ where the process, starting at the origin
and staying below the level $M$, arrives at $M-\epsilon_1$
(${\rm II}$) starting at $M-\epsilon_1$ at time $t_{\rm m}$ the
process arrives at $\epsilon_2>0$ at time $t_{\rm l}$, while 
staying below $M$ and (${\rm III}$) starting
at $\epsilon_2$ the process stays below $M$ and does not cross the origin
during $[t_{\rm l}, t]$. For each of these intervals, we will compute
the propagator of the path and then take their product to compute
the full propagator satisfying all the constraints.

\vspace{0.3cm}

\noindent{\bf Interval ${\rm I}$ ($0\le \tau\le t_{\rm m}$).} In this case, we have
a Brownian path propagating from $0$ at $\tau=0$ to $x(t_{\rm m})= M-\epsilon_1$
at time $\tau=t_{\rm m}$, but with the restriction that the path stays below $M$.
Let $G(x,x_0,\tau|M)$ denotes the probability density of arriving at
$x$ at time $\tau$, starting from $x_0$ at $\tau=0$ and staying
below $M$ till time $\tau$. Then this propagator $G(x,x_0,\tau|M)$
satisfies the diffusion equation
\begin{equation}
\frac{\partial G}{\partial t}= D\, \frac{\partial^2 G}{\partial x^2}
\label{diff_M.1}
\end{equation}
in the region $x\le M$, with the absorbing boundary condition 
$G(x=M,x_0,\tau|M)=0$ and
the initial condition $G(x,x_0,\tau=0|M)= \delta(x-x_0)$. The solution can
be easily found using the method of images~\cite{BF2005} and it reads
\begin{multline}
  G(x,x_0,\tau|M)= \\
  \frac{1}{\sqrt{4\, \pi\, D\, \tau}}\, \left[ e^{-(x-x_0)^2/{4D\tau}}
- \change{e^{-(2M-x-x_0)^2/{4D\tau}}}\right]\, .
\label{image.1}
\end{multline}

Setting $x= M-\epsilon_1$, $x_0=0$, $\tau=t_{\rm m}$ and expanding up to $O(\epsilon_1)$,
we get the propagator $G_I$ for part I
\begin{equation}
G_{\rm I}\approx \frac{\epsilon_1}{D}\, \frac{M}{\sqrt{4\, \pi\, D\, t_{\rm m}^3}}\, 
e^{-M^2/{4D t_{\rm m}}} \, .
\label{GI.1}
\end{equation}

\vspace{0.3cm}

\noindent{\bf Interval ${\rm II}$ ($t_{\rm m}\le \tau\le t_{\rm l}$).} In this second interval,
the path starts at $M-\epsilon_1$ and reaches $\epsilon_2$ at time $t_{\rm l}$,
while staying below $M$ during the trip. Hence, we can again use the
general result in Eq. (\ref{image.1}) and set $x_0=M-\epsilon_1$,
$x=\epsilon_2$, $\tau= t_{\rm l}-t_{\rm m}$ to get the propagator $G_{\rm II}$
for the interval $II$. In addition, expanding in powers of $\epsilon_1$
and $\epsilon_2$, the leading nonzero term is given by
\begin{equation}
G_{\rm II}\approx \frac{\epsilon_1\, \epsilon_2}{D}\, 
\frac{M}{\sqrt{4\, \pi\, D\, (t_{\rm l}-t_{\rm m})^3}}\, e^{-M^2/{4D (t_{\rm l}-t_{\rm m})}}\, .
\label{GII.1}
\end{equation}

\vspace{0.3cm}

\noindent{\bf Interval ${\rm III}$ ($t_{\rm l}\le \tau\le t$).}
In this interval, the path propagates during
the interval $\tau\in [t_{\rm l},t]$, starting from $\epsilon_2$
at $\tau=t_{\rm l}$, and does not cross zero during the interval $[t_{\rm l},t]$.
In addition, the path stays below the level $M$ during
$[t_{\rm l},t]$.
There are now two possibilities: (a) $\epsilon_2>0$, in which case
the process stays positive during $[t_{\rm l},t]$ and (b) $\epsilon_2<0$, in
which case the path stays negative during $[t_{\rm l},t]$.
Let us first consider the case $(a)$ where $\epsilon_2>0$ and the
path stays inside the box $[0,M]$ during $[t_{\rm l},t]$, starting at
$\epsilon_2>0$ at the begining of the interval. Let
us then consider the Brownian propagator $G_{\rm box}(x, x_0, T|M)$
as the probability density for a Brownian path to arrive at 
$0\le x\le M$ at time
$T$, starting from $0\le x_0\le M$ at $T=0$, while staying
inside the box $[0,M]$ during $T$. This propagator again
satisfies the diffusion equation in $0\le x\le M$
\begin{equation}
\frac{\partial G_{\rm box}}{\partial T}= D\, \frac{\partial^2 G_{\rm box}}{\partial x^2}
\label{diff_box.1}
\end{equation}
with absorbing boundary conditions: $G_{\rm box}(x=M,x_0,T|M)=
G_{\rm box}(x=0, x_0, T|M)=0$. The exact solution can be easily
found and it reads
\begin{multline}
  G_{\rm box}(x,x_0,T|M)= \\
  \frac{2}{M}\, \sum_{n=1}^{\infty}
\sin\left(\frac{n\pi x}{M}\right)\,
\sin\left(\frac{n\pi x_0}{M}\right)\, e^{- n^2 \pi^2 D T/M^2}\, .
\label{Gbox.1}
\end{multline}
Integrating over the final position gives the survival probability
up to time $T$, starting at $x_0$,
\begin{multline}
  S(x_0,T|M)= \\
  \frac{4}{\pi}\, \sum_{m=0}^{\infty} \frac{1}{2m+1}\,
\sin\left( \frac{(2m+1) \pi x_0}{M}\right)\, e^{-(2m+1)^2 \pi^2 D T/M^2}\,.
\label{Gsurv.1}
\end{multline}

Setting $T= t-t_{\rm l}$, $x_0= \epsilon_2$ in Eq. (\ref{Gsurv.1})
and subsequently taking the limit $\epsilon_2\to 0$
then gives the propagator for the interval ${\rm III}$ in the case (a)
when $\epsilon_2>0$. We get, to leading order in $\epsilon_2$,
\begin{equation}
G_{\rm III}^{(a)}\approx \frac{4\epsilon_2\, \theta(\epsilon_2)}{M}\,
\sum_{m=0}^{\infty} e^{-(2m+1)^2 \pi^2 D (t-t_{\rm l})/M^2}\, .
\label{GIIIa.1}
\end{equation}

Next we consider the case (b) where $\epsilon_2<0$. In this case,
the walk starting at $\epsilon_2<0$, stays negative during $[t_{\rm l},t]$
(the fact that it stays negative automatically guarantees that it
stays below the level $M$ since $M\ge 0$). The
probability density $G_{-}(x,x_0,T)$ for a Brownian motion to reach $x<0$, 
starting at $x_0<0$, and staying below $0$ during an interval $T$,
can again be computed by the method of images
and is given simply by~\cite{BF2005}
\begin{multline}
  G_{-}(x,x_0,T)= \\
  \frac{1}{\sqrt{4\, \pi\, D\, T}}\, 
\left[e^{-(x-x_0)^2/{4 D T}}- e^{-(x+x_0)^2/{4 D T}}\right]\, .
\label{dens_IIIb.1}
\end{multline}
Integrating over $x\le 0$ gives the survival probability
\begin{multline}
  S_{-}(x_0,T)= {\rm erf}\left(\frac{|x_0|}{\sqrt{4\, D\, T}}\right)\,\\
  {\rm where}\quad {\rm erf}(z) = \frac{2}{\sqrt{\pi}}\, \int_0^{z} e^{-u^2}\, du\, .
\label{survb.1}
\end{multline}
Hence, setting $x_0= \epsilon_2<0$, $T=t-t_{\rm l}$ and expanding to leading
order in $\epsilon_2$ then gives the propagator for interval ${\rm III}$
for the case (b) where $\epsilon_2<0$
\begin{equation}
G_{\rm III}^{(b)}\approx \frac{\epsilon_2\, \theta(-\epsilon_2)}{\sqrt{ \pi\, 
D\, (t-t_{\rm l})}}\, .
\label{GIIIb.1}
\end{equation}

Finally, combining the two results in Eqs. (\ref{GIIIa.1}) for $\epsilon_2>0$
and (\ref{GIIIb.1}) for $\epsilon_2<0$, we get
\begin{multline}
  G_{\rm III}\approx\\
  |\epsilon_2|\, \left[\frac{4}{M}\, \sum_{m=0}^{\infty} 
e^{-(2m+1)^2 \pi^2 D (t-t_{\rm l})/M^2} + 
\frac{1}{\sqrt{\pi \, D\, (t-t_{\rm l})}}\right]\, . 
\label{GIII.1}
\end{multline}
It is sometimes convenient to express the sum in Eq. (\ref{GIII.1})
slightly differently using the following Poisson summation formula
\begin{equation}
\sum_{m=0}^{\infty} e^{- \frac{(2m+1)^2 \pi^2}{4 z}}= \sqrt{\frac{z}{4\pi}}\,
\left[1+ 2\, \sum_{n=1}^{\infty} (-1)^n e^{-n^2 z}\right]
\label{Poisson.1}
\end{equation}
for $z>0$.
Using this formula in Eq. (\ref{GIII.1}) gives us a more compact formula
\begin{equation}
G_{\rm III}\approx \frac{2\, |\epsilon_2|}{\sqrt{\pi\, D\, (t-t_{\rm l})}}\, 
\sum_{n=0}^{\infty} (-1)^n \, e^{-\frac{n^2 M^2}{4\, D\, (t-t_{\rm l})}}\, .
\label{GIII.2}
\end{equation}

Now that we have the required propagators for all the three intervals
${\rm I}$, ${\rm II}$ and ${\rm III}$ respectively
in Eqs. (\ref{GI.1}), (\ref{GII.1}) and (\ref{GIII.2}), we take
their product using the Markov property of the process and
obtain the joint distribution $P(t_{\rm l},t_{\rm m},M|t)$ for $t_{\rm l}>t_{\rm m}$ as
\begin{widetext}
\begin{equation}
P(t_{\rm l}, t_{\rm m}, M|t)= \left[\lim_{\epsilon_1\to 0}{\cal N}_1(\epsilon_1) 
\frac{\epsilon_1^2}{D}\right]\,
\left[\lim_{\epsilon_2\to 0}{\cal N}_2(\epsilon_2)\frac{\epsilon_2^2}{D}
\right]\,  
\frac{\theta(t_{\rm l}-t_{\rm m})\, M^2}{2\, (\pi\, D)^{3/2}\, \sqrt{t-t_{\rm l}}\,
t_{\rm m}^{3/2} (t_{\rm l}-t_{\rm m})^{3/2}}\, \sum_{n=0}^{\infty} (-1)^n\, 
e^{-\frac{n^2 M^2}{4\, D\, (t-t_{\rm l})}}\, ,
\label{propM1.1}
\end{equation}
\end{widetext}
where ${\cal N}_1(\epsilon_1)$ and ${\cal N}_2(\epsilon_2)$ are
proportionality constants that are fixed by the overall normalization
of the joint distribution. As in Eq. (\ref{norm_fix.1}) in the
previous section, one can show that for proper normalization
we must have the identities
\begin{equation}
\lim_{\epsilon_1\to 0} {\cal N}_1(\epsilon_1)\,
\frac{\epsilon_1^2}{D} = 1 \quad {\rm and}
\quad 
\lim_{\epsilon_2\to 0} {\cal N}_2(\epsilon_2)\,
\frac{\epsilon_2^2}{D} = 1 \, .
\label{norm_fix.2}
\end{equation}
Substituting these results in Eq. (\ref{propM1.1}) gives us 
a nice compact result for $t_{\rm l}>t_{\rm m}$
\begin{multline}
P(t_{\rm l}, t_{\rm m}, M|t)= \\
\frac{\theta(t_{\rm l}-t_{\rm m})\, M^2}{2\, (\pi\, D)^{3/2}\, \sqrt{t-t_{\rm l}}\,
t_{\rm m}^{3/2} (t_{\rm l}-t_{\rm m})^{3/2}}\, \sum_{n=0}^{\infty} (-1)^n\,
e^{-\frac{n^2 M^2}{4\, D\, (t-t_{\rm l})}}\, .
\label{propM1.2}
\end{multline}
Finally, the integral over $M$ can now be performed term by term
to obtain the joint distribution $P_{13}^{>}(t_{\rm l},t_{\rm m}|t)$ (with
the superscript signifying $t_{\rm l}>t_{\rm m}$) which reads
\begin{multline}
P_{13}^{>}(t_{\rm l},t_{\rm m}|t)= \frac{\theta(t_{\rm l}-t_{\rm m})}{\pi\, t_{\rm l}^{3/2}\,
\sqrt{t-t_{\rm l}}}\,
\sum_{n=0}^{\infty} \frac{(-1)^n}{(1+\alpha\, n^2)^{3/2}}\\
    {\rm where}\quad \alpha=
    \frac{t_{\rm m}\, (t_{\rm l}-t_{\rm m})}{t_{\rm l}\, (t-t_{\rm l})}\, .
\label{P13_gt.1}
\end{multline}

\vspace{0.3cm}

\begin{figure}
\centering
\includegraphics[width = 0.95 \linewidth]{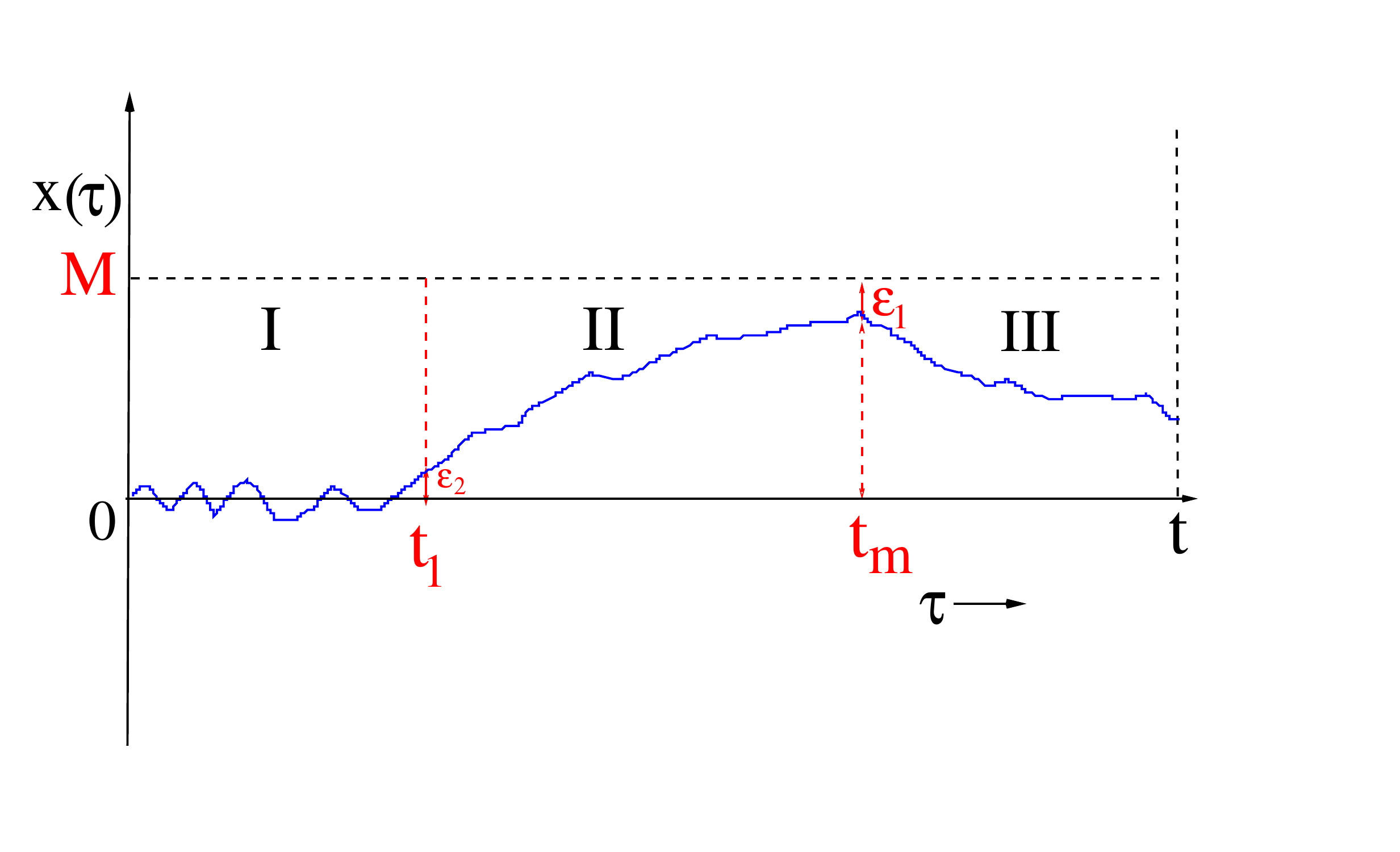}
\caption{\label{Fig13:traj2} A typical trajectory of a $1$-d
Brownian motion $x(\tau)$ of total duration $t$, starting at the
origin $x(0)=0$. The trajectory arrives at $\epsilon_2>0$
at time $t_{\rm l}$ while staying below $M$ and does not cross zero after time 
$t_{\rm l}$. It also
arrives at its maximum value $M-\epsilon_1$ at time $t_{\rm m}$.
We divide
the total duration $[0,t]$ into three intervals:
(${\rm I}$) the interval $[0,t_{\rm l}]$ where the process, starting at the origin
and staying below the level $M$, arrives at $\epsilon_2$
at time $t_{\rm l}$, (${\rm II}$) starting at $\epsilon_2$ at time $t_{\rm l}$ the
process arrives at $M-\epsilon_1$ at time $t_{\rm m}>t_{\rm l}$, while
staying below $M$ and also above $0$ and (${\rm III}$) starting
at $M-\epsilon_1$ at time $t_{\rm m}$, the process stays inside the
box $[0,M]$ during the interval $[t_{\rm m},t]$. 
Eventually, we take the limits $\epsilon_1\to 0$
and $\epsilon_2\to 0$ at the end of the calculation.}
\end{figure}

\noindent{\bf The case $t_{\rm l}<t_{\rm m}$.} In this case, the process starts from $0$
and arrives at $\epsilon_2>0$ at time $t_{\rm l}$, while staying below $M$--this
is region ${\rm I}$ in Fig. (\ref{Fig13:traj2}). Then, in region ${\rm II}$,
the process starting at $\epsilon_2$ at time $t_{\rm l}$, reaches at $M-\epsilon_1$
at time $t_{\rm m}$, while staying in the box $x\in [0,M]$ during the
time interval $[t_{\rm l}, t_{\rm m}]$. Finally in region ${\rm III}$, the process
starting at $M-\epsilon_1$ at time $t_{\rm m}$ stays inside the box
$x\in [0,M]$ during the rest of the time interval $[t_{\rm m},t]$.
As in the previos case, we will compute the propagators in
each of these three regions separately to leading orders
for small $\epsilon_1$ and $\epsilon_2$ and then take their
product to compute the joint distribution $P(t_{\rm l},t_{\rm m},M|t)$
with $t_{\rm l}<t_{\rm m}$.

\vspace{0.3cm}

\noindent{\bf Interval ${\rm I}$ ($0\le \tau\le t_{\rm l}$).} Here we can
directly use the result in Eq. (\ref{image.1}) obtained via the method of 
images with the substitution $x=\epsilon_2$, $x_0=0$ and $\tau=t_{\rm l}$.
This gives, after expanding in powers of $\epsilon_2$ and keeping only the
leading order term,
\begin{equation}
G_{\rm I}\approx \frac{\left[1- e^{-M^2/{ D\, t_{\rm l}}}\right]}{\sqrt{4\, 
\pi\, D\, t_{\rm l}}}\, \epsilon_2\, .
\label{GI2.1}
\end{equation}

\vspace{0.3cm}

\noindent{\bf Interval ${\rm II}$ ($t_{\rm l}\le \tau\le t_{\rm m}$).} In this
regime, the process moves in a box $x\in [0,M]$ starting from $\epsilon_2>0$
at time $t_{\rm l}$ and arriving at $M-\epsilon_2$ at time $t_{\rm m}>t_{\rm l}$. Hence
we can use the box propagator in Eq. (\ref{Gbox.1}) with
the substitution $x= M-\epsilon_1$, $x_0= \epsilon_2$ and $T= t_{\rm m}-t_{\rm l}$.
Expanding again for small $\epsilon_1$ and $\epsilon_2$, we get to
leading order
\begin{equation}
G_{\rm II}\approx \frac{2\pi^2 \epsilon_1\, \epsilon_2}{M^3}\,
\sum_{n=1}^{\infty} (-1)^{n-1}\, n^2\, e^{- n^2 \pi^2 D (t_{\rm m}-t_{\rm l})/M^2}\, .
\label{GII2.1}
\end{equation}

\vspace{0.3cm}

\noindent{\bf Interval ${\rm III}$ ($t_{\rm m}\le \tau\le t$).} 
In regime ${\rm III}$, the process again moves in the box $x\in [0,M]$,
starting at $M-\epsilon_1$ and surviving till time $t$, with absorbing boundary conditions at both $x=0$ and $x=M$. Hence, we can again use
the result for the survival probability $S(x_0,T|M)$ in Eq. (\ref{Gsurv.1})
with the substitution $x_0=M-\epsilon_1$ and $T=t-t_{\rm m}$. Exapanding it
for small $\epsilon_1$, one gets to leading order the propagator
for regime ${\rm III}$
\begin{equation}
G_{\rm III}\approx \frac{4 \epsilon_1}{M}\, \sum_{m=0}^{\infty}\,
e^{-(2m+1)^2 \pi^2 D(t-t_{\rm m})/M^2}\, .
\label{GIII2.1}
\end{equation}

Finally, taking the product of the three regimes,
respectively in Eqs. (\ref{GI2.1}), (\ref{GII2.1})
and (\ref{GIII2.1}), gives
the joint distribution $P(t_{\rm l},t_{\rm m},M|t)$ for
$t_{\rm l}<t_{\rm m}$ as
\begin{multline}
P(t_{\rm l}, t_{\rm m}, M|t)=
\frac{\theta(t_{\rm m}-t_{\rm l})\, 8 \pi^2\, D^2 \left(1- e^{-M^2/{Dt_{\rm l}}}\right)}
     {\sqrt{4\,\pi\, D\, t_{\rm l}}\, M^4}\times\\
\sum_{n,m=0}^{\infty} 
(-1)^{n-1} n^2\, e^{- \frac{\pi^2\, D}{M^2}\, \left( n^2(t_{\rm m}-t_{\rm l})
+ (2m+1)^2 (t-t_{\rm m})\right)}\, ,
\label{prop_full.1}
\end{multline}
where we used exactly the same normalization
as in Eq. (\ref{norm_fix.2}).
We now have to integrate over $M$. For this we use the following identity
\begin{equation}
\int_0^{\infty} \frac{1}{M^4}\, e^{-b\, M^2+a/M^2}\, dM = 
\frac{\sqrt{\pi} \left(1+2 \sqrt{ab}\right)}{4 a^{3/2}}\, e^{-2 \sqrt{ab}}
\label{iden.1}
\end{equation}
for $a>0,\,\, b>0$ . 
We now integrate Eq. (\ref{prop_full.1}) over $M$ using
the identitity (\ref{iden.1}) upon setting
$a= \pi^2 D \left( n^2(t_{\rm m}-t_{\rm l}) + (2m+1)^2 (t-t_{\rm m})\right)$
and $b=1/(D\, t_{\rm l})$. Simplifying, we get the joint distribution 
$P_{13}^{<}(t_{\rm l},t_{\rm m}|t)$ (with
the superscript signifying $t_{\rm l}<t_{\rm m}$) which reads
\begin{multline}
  P_{13}^{<}(t_{\rm l},t_{\rm m}|t)=
  \frac{\theta(t_{\rm m}-t_{\rm l})}{\pi\, t_{\rm l}^2}\times \\
\sum_{n,m=0}^{\infty} \frac{(-1)^{n-1}\, n^2}{\gamma_{n,m}^{3/2}}\,
\left[1- \left(1+ 2\, \pi\, \sqrt{\gamma_{n,m}}\, \right)\, e^{-2\, \pi\, 
\sqrt{\gamma_{n,m}}}\, \right]\, ,
\label{P13_lt.1}
\end{multline}
where
\begin{equation}
\gamma_{n,m}= \frac{\left[n^2(t_{\rm m}-t_{\rm l}) +(2m+1)^2 (t-t_{\rm m})\right]}{t_{\rm l}}\, .
\label{gamma_def}
\end{equation}
Adding Eqs. (\ref{P13_gt.1}) and (\ref{P13_lt.1}), we get our final
answer for the joint distribution $P_{13}(t_{\rm l},t_{\rm m}|t)
=P_{13}^{>}(t_{\rm l},t_{\rm m}|t)+ P_{13}^{<}(t_{\rm l},t_{\rm m}|t)$ that
results in \eqref{jointP13.1}

\begin{figure}
\includegraphics[width = 0.95\linewidth]{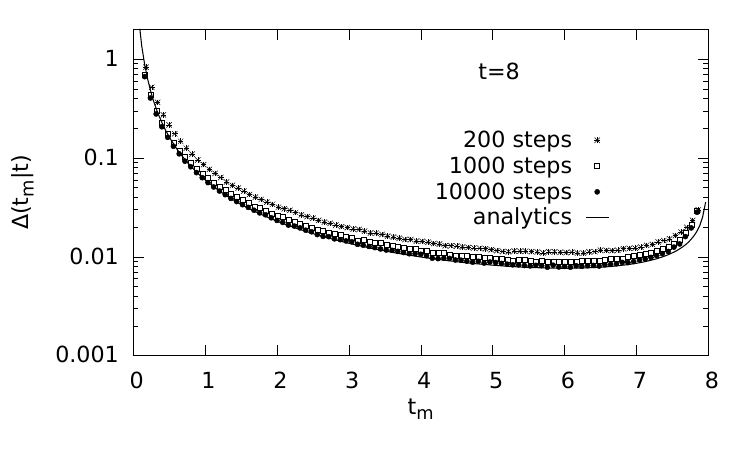}
\caption{The jump $\Delta(t_{\rm m},t)$ in the joint PDF joint $P(t_{\rm m},t_{\rm l}|t)$
as $t_{\rm l}\to t_{\rm m}^+$, plotted as a function of $t_{\rm m}$ for fixed $t=8$. The
symbols show simulation results with increasing number of steps
and the solid line represents the analytical result in Eq. (\ref{disc.1}).
The simulation results appraoch the analytical formula as the number of
steps increases.}
\label{fig:jumptm}
\end{figure}

\change{We have investigated the asymmetric behavior of the joint
  PDF for $t_{\rm l}=t_{\rm m}-\epsilon$ and
  $t_{\rm l}=t_{\rm m}+\epsilon$ as $\epsilon\to 0$.
  By analysing Eq. (\ref{jointP13.1}) we
find the following results as $t_{\rm l}$ approches $t_{\rm m}$ from above and
below,}
\begin{widetext}
\begin{eqnarray}
  P_{13}(t_{\rm l}, t_{\rm m}|t) \approx 
  \begin{cases}
&  \frac{1}{2\,\pi\, t_{\rm m}^{3/2}\, \sqrt{t-t_{\rm m}}} \, ,\quad\quad
\quad\quad\quad\quad\quad\quad\quad\quad\quad\quad\quad\quad\,\,
 {\rm as}\quad t_{\rm l}\to t_{\rm m}^+ \label{tmplus}\\
& \\
& C(t_{\rm m}, t)\, (t_{\rm m}-t_{\rm l})^{-7/4}\, \exp\left[-
\frac{\pi}{\sqrt{t_{\rm m}-t_{\rm l}}}\, \sqrt{\frac{t-t_{\rm m}}{t_{\rm m}}} \right]
\quad {\rm as} \quad t_{\rm l}\to t_{\rm m}^{-}\, ,
\end{cases}
\label{tmminus}
\end{eqnarray}
\end{widetext}
where the amplitude $C(t_{\rm m},t)$ can also be
computed explicitly and is given by
\begin{multline}
C(t_{\rm m},t)= \frac{1}{\sqrt{2}\, t_{\rm m}^2}\,
\left(\frac{t-t_{\rm m}}{t_{\rm m}}\right)^{1/4}\,\times \\
\left[1-\left[1+
2\, \pi\, \sqrt{\frac{t-t_{\rm m}}{t_{\rm m}}}\right]\,
e^{- 2\, \pi\, \sqrt{\frac{t-t_{\rm m}}{t_{\rm m}}}}\right]\, .
\label{C_def}
\end{multline}
Thus, as $t_{\rm l}\to t_{\rm m}$ from below, the joint distribution
vanishes extremely rapidly (with
an essential singularity
as seen in Eq. (\ref{tmminus})). In contrast, as $t_{\rm l}\to t_{\rm m}$ from above,
it approaches a nonzero constant given in Eq. (\ref{tmplus}).
Hence the discontinuity
\begin{eqnarray}
  \Delta(t_{\rm m},t) & = &
  P_{13}(t_{\rm l}=t_{\rm m}^+,t_{\rm m}|t)
  -P(t_{\rm l}=t_{\rm m}^{-},t_{\rm m}|t) \nonumber\\
& = & \frac{1}{2\,\pi\, t_{\rm m}^{3/2}\, \sqrt{t-t_{\rm m}}}\, ,
\label{disc.1}
\end{eqnarray}
is a non-monotonic function of $t_{\rm m}$ for fixed $t$.
In Fig. (\ref{fig:jumptm})
we plot $\Delta(t_{\rm m},t)$ as a function of $t_{\rm m}$ for fixed $t$ and
compare it to simulations.
The agreement between the analytical result in
Eq. (\ref{disc.1}) and the simulation
results is excellent.

\change{Finally, although in principle possible, computing
the joint moments of $t_{\rm l}$ and $t_{\rm m}$ from the joint distribution
seems a bit hard compared to the two previous cases. Thus, we did not perform
this task.}

\section{Feynman-Kac method to compute the distribution of the occupation time
for a bridge of duration $t$}

Consider a Brownian motion of duration $t_{\rm l}$, starting at the origin $x(0)=0$
and arriving at $x$ at time $t_{\rm l}$. We will finally set $x=0$ for our 
Brownian bridge. Let $t_{\rm o}= \int_0^{t_{\rm l}} \theta\left(x(\tau)\right)\, d\tau$
denote the occupation time for this bridge of duration $t_{\rm l}$. 
Let us denote the joint PDF of $t_{\rm o}$ and the final position $x$
at time $t_{\rm l}$ by $P(x, t_{\rm o}|t_{\rm l})$. Let us now take the Laplace
transform with respect to $t_{\rm o}$ and define
\begin{equation}
\tilde{P}_p (x|t_{\rm l})= \int_0^{\infty} dt_{\rm o}\, e^{-p\, t_{\rm o}}\,
P(x, t_{\rm o}|t_{\rm l})\, .
\label{lap.1}
\end{equation}
Note that by definition $t_{\rm o}\le t_{\rm l}$, hence the PDF 
$P(x, t_{\rm o}|t_{\rm l})$ must vanish for $t_{\rm o}\ge t_{\rm l}$, i.e.,
the upper limit of the integral in Eq. (\ref{lap.1}) is strictly $t_{\rm l}$.
Now, $\tilde{P}_p(x|t_{\rm l})$ satisfies the Feynman-Kac equation that
reads~\cite{BF2005}
\begin{equation}
\frac{\partial \tilde{P}_p}{\partial t_{\rm l}}= D\, \frac{\partial^2 \tilde{P}_p}{\partial x^2}
- p\, \theta(x)\, \tilde{P}_p(x|t_{\rm l})\, ,
\label{fk.1}
\end{equation}
with the initial condition $\tilde{P}_p(x|t_{\rm l}=0)= \delta(x)$. The last condition
comes from the fact that when $t_{\rm l}=0$, the final position $x$ must coincide with
the initial position (since the particle didn't evolve). Since initially the
particle is at the origin, clearly then $\tilde{P}_p(x|t_{\rm l}=0)= \delta(x)$.

To solve the partial differential equation (PDE) (\ref{fk.1}), it is 
convenient further to take a Laplace transform with respect to $t_{\rm l}$ and define
\begin{equation}
R_p(x,s)= \int_0^{\infty} dt_{\rm l}\, e^{-s\, t_{\rm l}}\, \tilde{P}_p(x|t_{\rm l})\, .
\label{fk.2}
\end{equation}
Taking Laplace transform of Eq. (\ref{fk.1}) with respect to $t_{\rm l}$ and using
the initial condition, we obtain an ordinary second order differential equation
for $R_p(x,s)$
\begin{equation}
D\, \frac{d^2 R_p}{dx^2} - \left(s+ p\, \theta(x)\right)\, R_p=-\delta(x)\, ,
\label{fk_ode.1}
\end{equation}
subject to the boundary conditions: $R_p(x\to \pm \infty, s)=0$.
To solve this differential equation we solve it for $x<0$ and $x>0$ separately
and then match the solution at $x=0$. For $x<0$, the solution is simply
$R_p(x,s)= A\, e^{\sqrt{s/D}\, x}$ where $A$ is a constant.  Note
that we have used the boundary condition $R_p(x\to -\infty, s)=0$
to discard the exponentially growing solution. Similarly, for $x>0$,
we get $P_p(x,s)= A\, e^{-\sqrt{(s+p)/D}\, x}$, where we used
$R_p(x\to  \infty, s)=0$ and the continuity of the solution at $x=0$.
Finally, the unknown constant $A$ can be fixed by noting the jump
(in the derivative at $x=0$) condition (follows by integrating
Eq. (\ref{fk_ode.1}) across $x=0$): $R_p'(x\to 0^+,s)- R_p'(x\to 0^{-},s)=-1/D$.
This gives 
\begin{equation}
A= \frac{1}{\sqrt{D}\, \left(\sqrt{s}+\sqrt{s+p}\right)}\, .
\label{value_A}
\end{equation}
This gives the complete solution $R_p(x,s)$ in the Laplace space. However, we are
interested only in the bridge case when $x=0$. Hence, we have
\begin{multline}
\int_0^{\infty} dt_{\rm l}\, e^{-s\, t_{\rm l}}\, \int_0^{t_{\rm l}} dt_{\rm o}\, 
e^{-p\, t_{\rm o}}\, P(x=0, t_{\rm o}|t_{\rm l})= \\
R_p(x=0,s)= A= 
\frac{1}{\sqrt{D}\, \left(\sqrt{s}+\sqrt{s+p}\right)}\, .
\label{double_lap.1}
\end{multline}
Fortunately, the double Laplace transform can be exactly inverted and we get for
the interval $I$
\begin{equation}
P(x=0, t_{\rm o}|t_{\rm l})= \frac{1}{\sqrt{4\, \pi\, D\, t_{\rm l}^3}}\, 
\theta\left(t_{\rm l}- t_{\rm o}\right)\, .
\label{jointI.1}
\end{equation}
Thus, interestingly the distribution is flat as a function of $t_{\rm o}$.
The explicit dependence on $t_{\rm o}$ is only in the upper bound (through
the theta function). We are not done yet. Because $P(x=0, t_{\rm o}|t_{\rm l})$
is the joint probability that the process in $[0,t_{\rm l}]$ has occupation time
$t_{\rm o}$ and that the end position of the process is at $x=0$.
To calculate the conditional PDF of $t_{\rm o}$ (given $t_{\rm l}$), we need
to divide it by the probability that the process reaches $x=0$ at time $t_{\rm l}$,
which is simply $1/\sqrt{4\, \pi\, D\, t_{\rm l}}$ for a Brownian motion of 
duration $t_{\rm l}$.
Hence, dividing Eq. (\ref{jointI.1}) by $1/\sqrt{4\, \pi\, D\, t_{\rm l}}$, we finally get
the PDF of $t_{\rm o}$, given $t_{\rm l}$, as simply a uniform distribution
\begin{equation}
p(t_{\rm o},t_{\rm l})= \frac{1}{t_{\rm l}}\, \theta\left(t_{\rm l}- t_{\rm o}\right)\, .
\label{pleft.2}
\end{equation}

\end{appendix}


\end{document}